\documentclass{aa}  

\usepackage{graphicx}
\usepackage{longtable}
\usepackage{soul}
\usepackage{txfonts}
\usepackage[colorlinks=true, allcolors=blue]{hyperref}
\usepackage[normalem]{ulem}
\usepackage{xcolor}
\DeclareUnicodeCharacter{0301}{\'{e}}
\DeclareUnicodeCharacter{0301}{\'{i}}
\DeclareUnicodeCharacter{2212}{-}

\usepackage{multirow}

\begin{document}

   \title{Spectral characterization of young LT dwarfs}
   \authorrunning{Piscarreta et al.}

   \author{L. Piscarreta
          \inst{1,2}, K. Mu\v{z}i\'c\inst{3}, V. Almendros-Abad\inst{4} and A. Scholz\inst{5}
          }

   \institute{CENTRA, Faculdade de Ci\^{e}ncias, Universidade de Lisboa, Ed. C8, Campo Grande, P-1749-016 Lisboa, Portugal\\
              \email{lara.alvopiscarreta@eso.org}
         \and
         European Southern Observatory, Karl-Schwarzschild-Strasse 2, 85748 Garching bei München, Germany
         \and
Instituto de Astrofísica e Ciências do Espaço, Faculdade de Ciências, Universidade de Lisboa, Ed. C8, Campo Grande, 1749-016 Lisbon, Portugal
        \and
            Istituto Nazionale di Astrofisica (INAF) – Osservatorio Astronomico di Palermo, Piazza del Parlamento 1, 90134 Palermo, Italy
        \and
             SUPA, School of Physics \& Astronomy, University of St Andrews, North Haugh, St Andrews, KY16 9SS, United Kingdom\\
             }

   \date{Received; accepted}

 
  \abstract
   {JWST and next-generation facilities are expected to uncover populations of free-floating objects below the deuterium-burning limit in a number of young clusters and star-forming regions. These young planetary-mass brown dwarfs have spectral types L and T, shaped by molecular absorption and modified by their low gravity, which makes them distinct from field objects.} 
   {We aim to provide a detailed characterization of near-infrared spectra for young LT brown dwarfs, including robust spectral typing, calibrating spectral indices, identifying possible binaries, and selecting suitable spectral standards.}
   {We process and analyze archival spectra from VLT/X-shooter for a sample of 56 dwarfs with ages between 10 and 600\,Myr and spectral types between late-M and mid-T. We re-determine spectral types by comparing them with a set of literature templates. We assess a large range of spectral indices, calibrated using a specifically designed literature sample.}
   {We identify 15 spectral indices that are useful for spectral typing for specific spectral ranges discussed here and provide the scaling relations with spectral types. We also identify 6 spectral indices which can be used to separate young L dwarfs from the field ones. The EWs of the alkali lines show a correlation with age, increasing towards the objects with higher surface gravity. From our sample, we confirm 3 that are likely to be binaries by their anomalous spectra that appear to be better fitted by a combination of spectral types. We identify 12 objects as preliminary near-infrared spectral standards for young LT dwarfs.}
   {This paper presents a significant step toward understanding the spectral sequence and properties of young L and T dwarfs. The relations and standards provided here will be useful for future spectroscopic work on young brown dwarfs and giant planets.}


   \maketitle
%

\section{Introduction}

The bottom of the initial mass function (IMF) is populated by brown dwarfs (BDs), substellar objects with masses $\lesssim80$\,M$_{Jup}$. The lightest among these are the so-called free-floating planetary-mass objects (PMOs) with masses below the deuterium-burning limit at $\sim12$\,M$_{Jup}$, whose existence have been confirmed by various deep surveys in nearby young star-forming regions \citep[SFRs;][]{Luhman_2009,scholz_2012,Pena_2012,Lodieu_2018,Miret-Roig_2022,Bouy_2022}. However, the details of the free-floating PMO formation process are still largely unknown and represent one of the important missing pieces in our understanding of star formation. Broadly speaking, populations of isolated PMOs in star clusters may form by cloud fragmentation and core collapse (star-like formation), or in protoplanetary disks (planet-like formation), followed by ejection due to encounters with other stars or planet-planet interactions (see \citealt{Miret-Roig_2022_review} and references therein). The relative importance of these two scenarios is not known, mainly due to the lack of observational constraints - given their intrinsic faintness, studies
of PMOs present a challenging task for most of the current facilities. In total, only about a few dozen young
isolated PMOs have been spectroscopically confirmed so far, with estimated masses above $\sim$5\,M$_{Jup}$. This situation is expected to change drastically in the near future with the advent of the James Webb Space Telescope (JWST), which will provide, for the first time, a robust census of young (1$-$3\,Myr) free-floating PMOs in the 1-15\,M$_{Jup}$ range \citep{Scholz_2022, Pearson2023}. 
These objects are expected to have spectral types (SpTs) in the range between $\sim$L0 and early-to-mid T-types. 
Evolutionary models predict that a Jupiter-mass object at 1-3 Myr should have an effective temperature in the range 700 - 900 K, showing the typical T-type features in the model atmospheres \citep{Sonora-bobcat}.

Spectral characteristics of BD atmospheres are strongly influenced by gravity, which has been extensively used to remove field contaminants from spectroscopic samples of BD candidates in star-forming regions (\citealt{Bayo_2011, Muzic_2015, Esplin_2020}, to name only a few). In the near-infrared (NIR), gravity-sensitive features include the alkali lines (K I, Na I) and FeH absorption bands, which are less pronounced in young, low-gravity atmospheres, as well as the broadband form of the H-band, shaped by water absorption, which appears sharply triangular in young atmospheres as opposed to a rounder shape in field dwarf atmospheres \citep{Gorlova_2003,McGovern_2004,Allers_Liu_2013,Manjavacas_2020,AlmendrosAbad_2022}. The assessment of youth (and therefore the membership in a cluster or a SFR) and the SpT estimation are typically performed through a comparison of candidate spectra with a set of spectral templates or with the help of numerous spectral indices derived for both purposes (see \citealt{AlmendrosAbad_2022} and references therein). But, while the LT-type field sequence has been robustly defined \citep{Kirkpatrick_2010,Burgasser_2006}, high-quality spectra of young objects are rare, making
the existing young set of templates sparse for the L-types \citep{Luhman_2017}, and basically non-existent for the T-types. On the other hand, L-type field dwarfs  exhibit significantly bluer colors than their young counterparts \citep{Faherty_2016,Schneider_2017}, making them unsuitable for spectral type derivation of young objects. 

The current sample of spectroscopically confirmed L-type members in nearby SFRs and young clusters with ages $\lesssim 5\,$Myr, apart from being small, extends only down to SpTs L3-L4 \citep[e.g.][]{Alvesdeoliveira_2012, Muzic_2015}, and typically suffer from significant extinction, which may introduce additional uncertainties in the spectral fitting process. For both their vicinity and lack of extinction, it is useful to turn to nearby young moving groups (NYMGs) in order to search for a sample of relatively young (10$-$600\,Myr) objects that would span a large range of SpTs. Although due to differences in age not all the LT objects in NYMGs will be in the planetary-mass regime, they are still expected to share various spectral properties with the younger objects at the same effective temperature. 
This kind of sample can serve for the comparison with JWST spectra of young free-floating PMO candidates, as well as with directly imaged giant planetary-mass companions \citep{Faherty_2016}. 

This paper builds upon the analysis presented in \citet{AlmendrosAbad_2022}, who analyzed a large sample of NIR spectra of young and old cool dwarfs in the spectral range of M0 to $\sim$L3. They inspected a number of SpT- and gravity-sensitive spectral indices available in the literature and defined several new ones. Using machine learning models, they confirmed that the broadband shape of the $H-$band represents the most relevant feature when it comes to distinguishing low- from high-gravity atmospheres, followed by the FeH absorption bands and alkali lines in the $J-$band portion of the spectrum.
The main goal of this work is to extend the analysis of \citet{AlmendrosAbad_2022} to later spectral types and provide "prescriptions" to obtain an accurate SpT and gravity classification for mid-to-late L dwarfs, and eventually the T-dwarfs. 
To that end, we present a collection of NIR spectra of LT members of various NYMGs, SFRs, and clusters with ages $\gtrsim10$\,Myr. This is a starting point towards constructing a more complete set of templates of young LT dwarfs.

There are several other works in the literature that also deal with classification and youth assessment of the solar-neighbourhood L- \citep{Allers_Liu_2013, Gagne_2015_VII,Martin_2017,Cruz_2018} and T-dwarfs \citep{Zhang_2021}. However, the spectral-type coverage  of the proposed templates extends down to $\sim$L4, and is quite patchy at later SpTs. Any new spectra are therefore helpful to fill-in the gaps. A large fraction of the above mentioned spectral compendium come from the northern facilities, it is then useful to turn to the South, where the sky is particularly rich in NYMG members. Furthermore, a large fraction of the available spectra have either very-low spectral resolution (R of a few hundred), or a limited wavelength coverage (e.g. the $J$-band in \citealt{Martin_2017}). A wide spectral coverage in the NIR is important for obtaining a proper assessment of the extinction which is fundamental for the characterization of new planetary-mass members of SFRs. At medium spectral resolutions, atomic and molecular features are more easily separated, which in the first place provides us with better line width measurements, and even increases the number of lines that can be used for youth assessment. Also,
medium-spectral resolution enables an in-depth analysis of the objects' atmospheric parameters (e.g. effective temperature, gravity, metallicity), along with the assessment of their cloud properties and disequilibrium chemistry \citep{Petrus_2023, Hood_2023}. Although a retrieval-type analysis is out of the scope of this paper, another goal of this work is to provide spectra as the basis for future studies that will help
improve the brown dwarf and planetary atmosphere models.

In Sect.~\ref{sec:data}, we present our main dataset and complementary ones, along with the details of the reduction process. We investigate the objects' position in a color-magnitude diagram (Sect.~\ref{sec:cmd}), derive their SpT and extinction (Sect.~\ref{sec:spt_ext}), and evaluate various features related to gravity classification (Sect.~\ref{sec:youth}). 
In Sect.~\ref{sect:standards}, we define a preliminary list of objects that can be used as standards for spectral typing of young L and T dwarfs. The summary and conclusions are presented in Sect.~\ref{sec:summary}.

\section{Dataset and data reduction}
\label{sec:data}
\subsection{Sample selection}\label{subsec:sample_selection}

Young PMOs with ages of $\sim$1$-$5\,Myr and masses $\sim$1$-$15\,M$_{Jup}$ should have spectral classifications equal to $\sim$L0 or later \citep{Scholz_2022}. With this in mind, and in order to be complete, the selection criteria for the dataset used in this work are objects located in NYMGs, nearby SFRs and clusters, with NIR SpTs equal to M8 or later and with publicly available spectra from the X-shooter spectrograph at ESO's Very Large Telescope (VLT). 
The sources were chosen from the list of UltraCoolDwarfs\footnote{\href{https://jgagneastro.com/list-of-ultracool-dwarfs/}{Jonathan Gagné's List of Ultracool Dwarfs}}, as well as from \citet{Allers_Liu_2013,Best_2015,Burgasser_2016,Kellog_2016,Manjavacas_2020,Marocco_2015,Naud_2014,Schneider_2016} and their NIR spectral classifications from the literature are between M8 and T5.5. 
We find that 56 objects match the described criteria. Of these, 34 objects belong to NYMGs ($\beta$ Pictoris Moving Group, BPMG; AB Doradus Moving Group, ABDMG; Argus, ARG; Carina-Near, CARN; Castor, CAS; Columba, COL; Tucana-Horologium, THA; and TW Hydrae, TWA), and 22 objects to nearby SFRs and clusters (Upper Scorpius, USCO; 32 Orionis, 32OR; Praesepe, PRA; and Pleiades, PLE). Table \ref{tab:xshoo_observations} summarizes the details of our sample and Table \ref{tab:NYMG_dist_age} the information regarding the mentioned regions.

\begin{table}
\caption{NYMGs, young clusters, and SFRs included in this work.}
\begin{center}
\resizebox{0.42\textwidth}{!}{
\begin{tabular}{lllll}
\hline\hline
Region & Distance (pc) & Ref & Age (Myr) & Ref \\ \hline
ABDMG & $30^{+20}_{-10}$ & (1) & 149$^{+51}_{-19}$ & (4) \\
ARG & 72.4 & (2) & 40$-$50 & (2) \\
BPMG & 30$^{+20}_{-10}$ & (1) & 24$\pm$3 & (4) \\
CARN & 30$\pm$20 & (1) & 200$\pm$50 & (5) \\
CAS & 5$-$20 & (9) & 320$\pm$80 & (9) \\
COL & 50$\pm$20 & (1) & 42$^{+6}_ {-4}$ & (4) \\
THA & $46^{+8}_{-6}$ & (1) & 45$\pm$4 & (4)\\
TWA & 60$\pm$10 & (1) & 10$\pm$3 & (4) \\
32OR & 96$\pm$2 & (1) & $22^{+4}_{-3}$ & (4) \\
PLE & 134$\pm$9 & (1) & 112$\pm$5 & (6) \\
PRA & 187.35$\pm$3.89 & (8) & $\sim$600 & (3) \\
USCO & 145$\pm$9 & (10) & 10$\pm$3 & (7)\\
\hline
\end{tabular}}
\end{center}
\tablefoot{(1) \citet{Gagne_2018_BANYAN}; (2) \citet{Zuckerman_2019}; (3) \citet{Kraus_2007}; (4) \citet{bell_2015}; (5) \citet{Zuckerman_2006}; (6) \citet{Dahm_2015}; (7) \citet{Pecaut_2016}; (8) \citet{Lodieu_2019}; (9) \citet{Ribas_2003} ; (10) This work.}
\label{tab:NYMG_dist_age}
\end{table}

In the analysis presented in the following sections, we divide the sample into four different classes based on age:

\begin{itemize}
    \item 10\,Myr $<$ Age $\leq$ 30\,Myr: objects belonging to USCO, TWA, BPMG, and 32OR;
    \item 30\,Myr $<$ Age $\leq$ 100\,Myr: objects belonging to ARG, COL, and THA;
    \item 100\,Myr $<$ Age $\lesssim$ 300\,Myr: objects belonging to ABDMG, CARN, CAS, and PLE. All the objects in this age class have ages between 100 and 200\,Myr except object \#27 which is considered a member of the Castor Moving Group ($\sim$320\,Myr) \footnote{The age of the Castor Moving Group (CAS) is not that well constrained and shows a large spread among its original and probable members \citep{Brandt_2014,Mamajek_2013}. But most of the stars in CAS seem to be older than 200\,Myr \citep{Zuckerman_2013}. In fact, it will be noticeable in the posterior analysis that the object \#27 has an age between 200\,Myr and the oldest age class considered (PRA, age $\sim$600\,Myr).};
    \item Age $\sim$ 600\,Myr: objects belonging to PRA.
\end{itemize}

\subsection{Data reduction}

X-shooter \citep{Vernet_2011} is an intermediate resolution spectrograph that covers a wide wavelength range from $\sim$300 to 2500\,nm, simultaneously, with its three independent arms: UVB arm (300-550\,nm), VIS arm (550-1000\,nm), and NIR arm (1000-2480\,nm). The latter is of the utmost relevance for this study, given the cool nature of the objects in our sample.

The publicly available data taken with X-shooter have been downloaded from the ESO archive, along with the appropriate calibrations. The EsoReflex environment \citep{esoreflex} was used to run version 3.5.3 of the X-shooter pipeline \citep{xshooter_pipeline} for the data reduction. The data reduction steps encompass dark current subtraction, flat-field correction, order tracing, wavelength calibration, sky subtraction, and flux calibration. The final one-dimensional spectrum was extracted from the rectified and order-merged 2D data product obtained from the pipeline. For that, we used the task \texttt{apall}, implementing the optimal spectral extraction algorithm, from IRAF\footnote{IRAF is distributed by the National Optical Astronomy Observatories, which are operated by the Association of Universities for Research in Astronomy, Inc., under cooperative agreement with the National Science Foundation.} (Image Reduction and Analysis Facility, \citealt{IRAF}).
The noise of the final 1D spectrum is extracted with the help of the 2D uncertainty map supplied by the pipeline.
We compared the IRAF extraction with the final 1D extracted spectra obtained from the X-shooter pipeline and found that for the majority of our sample, the IRAF extraction returned spectra with a better signal-to-noise ratio (SNR). 
The telluric correction was performed using the \textit{Molecfit} software \citep{Smette_2015,Kausch_2015}. This tool computes a synthetic transmission spectrum of the Earth's atmosphere considering the atmospheric conditions at the time of the observations.

For each object, the observing sequence typically provides several spectra, which were individually corrected. For the objects that have been observed on more than one night, we median-combined the telluric-corrected observations from each night. In each of these cases, we inspected whether the SNR of the combination improved that of an individual observation. If so, the final spectrum is the median-combined spectrum, if not, we keep the observation with a higher SNR.

In the NIR range, for spectra reduced and corrected for telluric absorption, the dominant noise contribution comes from OH line residuals from the sky subtraction process. We used the list of wavelengths corresponding to the sky emission lines from EsoReflex to remove the residuals of OH line subtraction above a defined threshold. The removed values were replaced by the mean flux in a narrow window around that wavelength. The wavelength ranges 1.138-1.143 $\mu$m, 1.165-1.185 $\mu$m, and 1.235-1.260 $\mu$m were not corrected in order not to interfere with the spectral lines present in this range: Na I doublet at $\sim$1.14 $\mu$m and the K I doublets at 1.169, 1.177 $\mu$m and 1.243, 1.252 $\mu$m \citep{Allers_Liu_2013}, respectively.

We note that some spectra seem to present a slight discontinuity around $\sim$2.27\,$\mu$m, which coincides with one of the CO band-heads. This is probably due to an illumination (vignetting) problem at the edge of the X-shooter 11th order \citep{Gonneau_2016, Lodieu_2018}.  

\subsection{Complementary spectroscopic dataset}

In several parts of the analysis presented in this paper, we use the sample of approximately 2760 publicly available NIR spectra of cool dwarfs compiled by
\citet{AlmendrosAbad_2022} (hereafter AA22). The spectra span SpTs between M0 and L5, and were obtained with a variety of telescopes and instruments. This dataset is divided according to age into the following classes: young ($\lesssim10\,$Myr), mid-gravity (objects belonging to NYMGs older than 10\,Myr, or those classified as INT-G by \citealt{Allers_Liu_2013}), and field objects (older objects with no youth spectroscopic features). The AA22 dataset was complemented by the objects with the literature SpT later than L5 from the SpeX Prism Spectral Library\footnote{\href{https://cass.ucsd.edu/~ajb/browndwarfs/spexprism/index.html}{SpeX Prism Library}} and the Montreal Spectral Library\footnote{\href{https://jgagneastro.com/the-montreal-spectral-library/}{Montreal Spectral Library}}. In total, 98 spectra were included. To have a homogeneous spectral classification for the entire complementary sample, we re-derived the SpTs for the added objects using the same method as in 
AA22, considering the T-type field templates from \citet{Burgasser_2006}. Of the 98 objects, 18 belong to the mid-gravity age class from AA22, whereas the remaining is inserted in the field class.
In several figures presented in this paper, we maintain the same color coding for the three age classes as in AA22: orange for the young objects, blue for the mid-gravity, and gray for the field.

\section{Color -- absolute magnitude diagram}
\label{sec:cmd}

We start our analysis by examining the properties of our sample in the color-absolute magnitude diagram. The magnitudes in the J-, H- and K-bands were retrieved from 2MASS \citep{Cutri_2003} and complemented with catalogs from several other authors \citep{Luhman_2020,Bouy_2022,Miret-Roig_2022,Medan_2021,Faherty_2016,Smart_2017,Burgasser_2016,Liu_2016,Marocco_2015,Lodieu_2018,Bouy_2015}. The parallaxes are from \textit{Gaia} EDR3 \citep{GaiaEDR3_1,GaiaEDR3_2} considering a matching radius of 2$''$ and are also complemented with measurements from several different authors \citep{Liu_2016,Best_2021,Galli_2017,Manjavacas_2019}. Only 28 of the initial 56 objects have trigonometric parallax measurements. For objects located in USCO and PRA that do not have a parallax measurement, we assumed the distance to be equal to the distance to the respective young association. We calculate the mean distance to USCO using \textit{Gaia} DR3 parallaxes and the latest literature census of members of this region \citep{Luhman_2022}. We find a mean distance to USCO of 145$\pm$9 pc (see Table \ref{tab:NYMG_dist_age}) which is in agreement with previous works (145$\pm$2pc, \citealt{Cook_2017}; 130$\pm$20pc, \citealt{Gagne_2018_BANYAN}). Object \#63 is the only one with a parallax measurement but no available 2MASS photometry. We implemented the magnitude system conversions from \citet{Leggett_2006} to convert the MKO magnitudes \citep{Zhang_2021} to 2MASS.

\begin{figure}
\centering
    \includegraphics[width=0.49\textwidth]{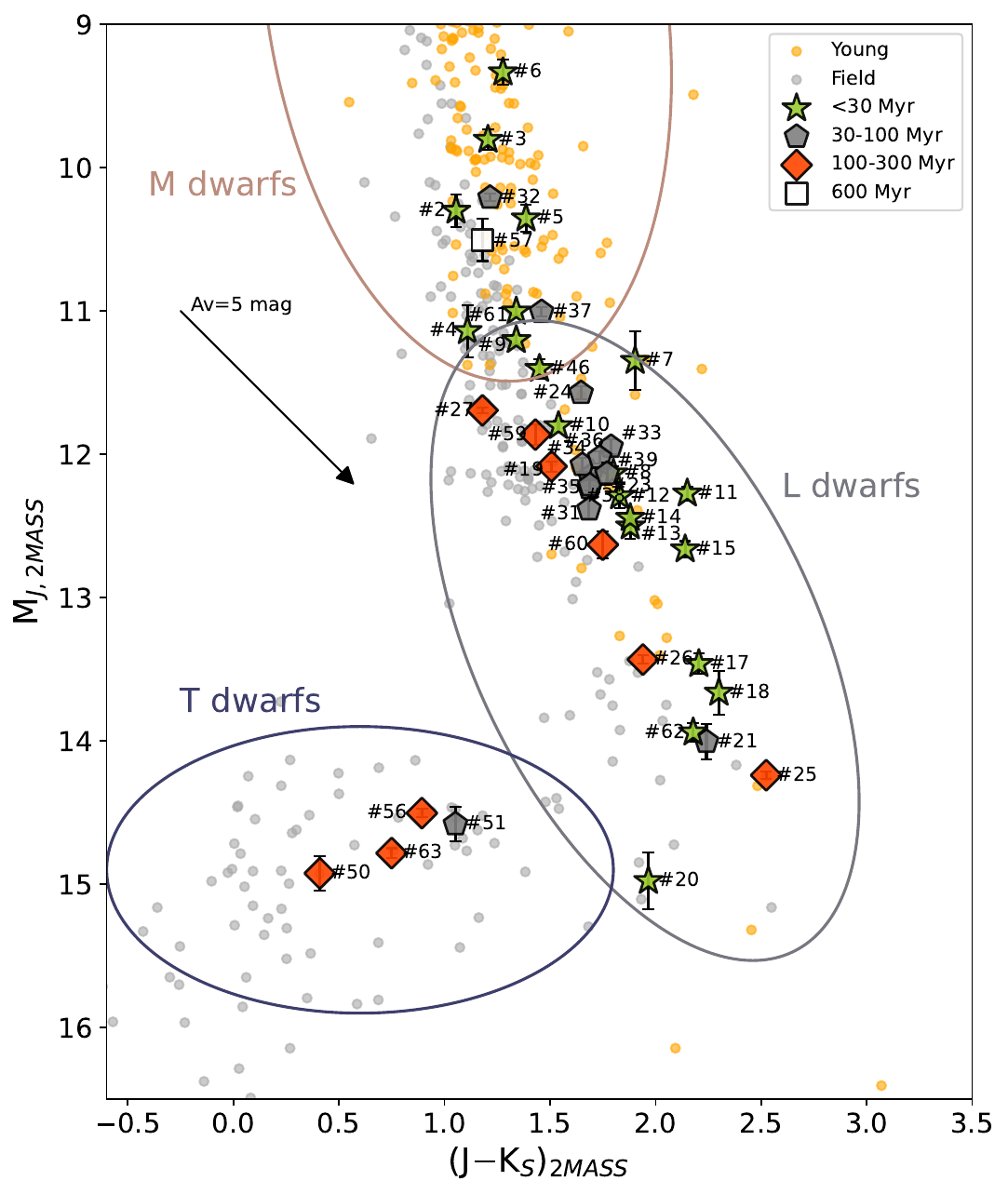}
\caption{(M$_{J}$, J-K$_{S}$) color-absolute magnitude diagram of our objects, very-low gravity (orange circles), and field objects (gray circles) from the literature (AA22; Database of Ultracool Parallaxes). Our sample is divided according to the age of the association to which the objects probably belong. Green stars represent objects in regions with ages $<$30\,Myr, dark gray pentagons with ages between 30$-$100\,Myr (included), red diamonds with ages between 100$-$300\,Myr (included), and white squares with ages $\sim$600\,Myr. Data points where the error bar is not visible mean that the bar is smaller than the marker size. The three ellipses approximately highlight where the M, L, and T dwarfs are located in this diagram. The arrow indicates what would be the effect of an addition of 5 magnitudes of extinction.}
\label{fig:cmd}
\end{figure}

Fig.~\ref{fig:cmd} exhibits the 2MASS absolute magnitude in the J-band as a function of (J-K$_{S}$) for 44 of our initial 56 objects. We overplotted young (ages $\lesssim$ 10\,Myr) and field objects from AA22 and L dwarfs, with gravity classifications of VL-G (very low-gravity) and FLD-G (field gravity), and T dwarfs from the Database of Ultracool Parallaxes\footnote{\url{http://www.as.utexas.edu/~tdupuy/plx/Database_of_Ultracool_Parallaxes.html}} \citep{Dupuy_2012,Liu_2016,Dupuy_2013}. The objects belonging to the young class of AA22 and those with gravity classification equal to VL-G are shown as orange dots whereas the FLD-G L dwarfs and T dwarfs as gray dots. The information regarding magnitudes and parallaxes for the objects in AA22 were also retrieved from 2MASS and \textit{Gaia} EDR3, respectively, considering the same matching radius as for our sample. For the case of the remaining objects, their 2MASS absolute magnitudes in the J-band and their (J-K$_{S}$) colors are provided in the respective paper. Contrary to the objects in our sample, these are corrected for extinction.  

The NIR colors of M- and L-type dwarfs are directly linked to their surface temperature: L dwarfs have a lower surface temperature thus they will exhibit redder colors. Moreover, young L dwarfs are typically redder than their field counterparts of the same spectral type \citep{Kirkpatrick_2008,Cruz_2009,Faherty_2016}, thought to be a consequence of their lower surface gravity leading to the formation of clouds in the upper layers of their atmospheres \citep{Madhusudhan_2011}, or, alternatively, of thermo-chemical instabilities in cloudless atmospheres \citep{Tremblin_2016}. As we move towards the lower temperatures, into the T-dwarf sequence, we can observe a dramatic shift towards bluer $J-K_{S}$ colors, often interpreted as a transition towards "cloudless" spectra, i.e. the clouds in these objects are expected to form so deep in the atmosphere that they have little or no influence on the final spectrum \citep{Faherty_2016}.

The ellipses shown in Fig.~\ref{fig:cmd} mark a rough location of different spectral classes. We can observe that the majority of the objects with expected SpT classifications of M and L appear to follow the young sequence. Nonetheless, some objects appear to be more in agreement with the field sequence (e.g, objects \#2, \#4, \#27, \#59, to name a few), which might suggest older ages.

\section{Spectral type and extinction}
\label{sec:spt_ext}
In this section, we derive the spectral types for the objects in our sample. To that end, we employ a comparison of the X-shooter spectra to a set of spectral templates (Sect.~\ref{subsec:spt_temp}), as well as a variety of SpT-sensitive indices (Sect.~\ref{subsec:spt_indices}). We also perform a methodical search for potential unresolved binaries in our sample (Sect.~\ref{subsec:binaries}).

\subsection{Comparison to spectral templates}\label{subsec:spt_temp}

The first method we use to derive the SpT is a direct comparison of the spectra with both young and field NIR spectral templates defined in the literature. The young spectral standards are from \citet{Luhman_2017}, whereas the field dwarf templates come from \citet{Kirkpatrick_2010} and \citet{Burgasser_2006}.
The L-type spectral sequence \citep{Luhman_2017} has been constructed using the solar-neighborhood objects with an age of $\lesssim$100 Myr, a subset of which have been associated with NYMGs.
Table \ref{tab:spectral_templates} summarizes the different sets of templates, along with the SpT ranges for which they were defined. The regions affected by the telluric absorption were excluded from the fitting process (1.3-1.5$\,\mu$m and 1.78-2.0$\,\mu$m), and  
the spectra were resampled to have the same wavelength grid of the template that they were being compared to.

\renewcommand{\arraystretch}{1.5}
\begin{table}
\centering
\caption{Sets of spectral templates used for the NIR SpT derivation.}
\resizebox{0.49\textwidth}{!}{
\begin{tabular}{lcc}
\hline
\hline
Type & SpT Range & Ref \\ \hline
Young & \begin{tabular}[c]{@{}c@{}}M0 - L0 at each 0.5 subclass interval\\ L2, L4 and L7 \end{tabular} & \citet{Luhman_2017} \\ \hline
Field & M0 - L9 at each 1 subclass interval & \citet{Kirkpatrick_2010} \\ \hline
Field & T0 - T8 at each 1 subclass interval & \citet{Burgasser_2006} \\ \hline
\end{tabular}}
\label{tab:spectral_templates}
\end{table}

In the fitting process, extinction may also be used as an additional free parameter. Within our sample, this is only relevant for the objects belonging to USCO, which suffer low, but non-negligible, amount of interstellar reddening \citep{Ardila_2000,Rizzuto_2015}. In this case, the extinction A$_{V}$ was varied between 0 to 2\,mag \citep{Lodieu_2008}, with a step of 0.2\,mag. We used the extinction law from \citet{Cardelli_1989} with R$_{V}$=3.1 to redden the spectral templates. 
For the remaining objects belonging to NYMGs (located within the so-called Local Bubble, d$\lesssim$100 pc) and the two open clusters (both PLE and PRA have low to negligible foreground reddening: $E(B-V)_ {PLE}=0.04$ mag, \citealt{Breger_1986}; $E(B-V)_ {PRA}=0.027\pm0.004$ mag, \citealt{Taylor_2006}), we set A$_{V}$=0 \citep{bell_2015,Pecaut_2013,Lodieu_2019}. 

Similarly to the method implemented in AA22, the fitting process has three variables - SpT, extinction, and normalization wavelength. The latter is searched on a grid from 1.55 to 1.78 $\mu$m with a step of 0.02 $\mu$m. For objects in associations other than USCO, there are only two variables (SpT and normalization wavelength). The best-fit template is the one that minimizes the $\chi^{2}$ parameter, defined as:
\begin{equation}
\centering
    \chi^{2} = \frac{1}{N - m} \sum\limits_{i=1}^{N} \frac{( O_{i} - T_{i} )^{2}}{\sigma_{i}^{2}},
\end{equation}
where $N$ is the number of points included in the fit (number of wavelength values), \textit{m} is the number of free parameters ($m=3$ for USCO and $m=2$ for the rest), $O$ and $\sigma$ are the object spectrum and its associated noise, respectively, and $T$ is the template spectrum. All spectra were individually compared with each one of the field and young templates. Next, we compare the $\chi^{2}$ values of both comparisons and use the template (young or field) associated with the lower $\chi^{2}$ value to derive the SpT and extinction, where applicable. As an estimate of the uncertainty on the derived SpT, we use the SpT step at which the correspondent set of templates is defined. In the case of young templates, if the SpT derived is L0 or L2 we assumed an uncertainty of $\pm$2.0 sub-SpTs, $^{+3.0}_{-2.0}$ sub-SpTs for L4 and $\pm$3.0 sub-SpTs for L7, due to the lack of young L1, L3, L5, L6, L8, and L9 young templates.

In several cases, the attributed best-fit SpT in the M- and L-type spectral range was that of a field template, which is somewhat unexpected given the youth of the sample. For these objects, we performed an additional check in order to visually inspect the quality of the best-fitted young and field templates. 

\begin{figure}
\centering
    \includegraphics[width=0.49\textwidth]{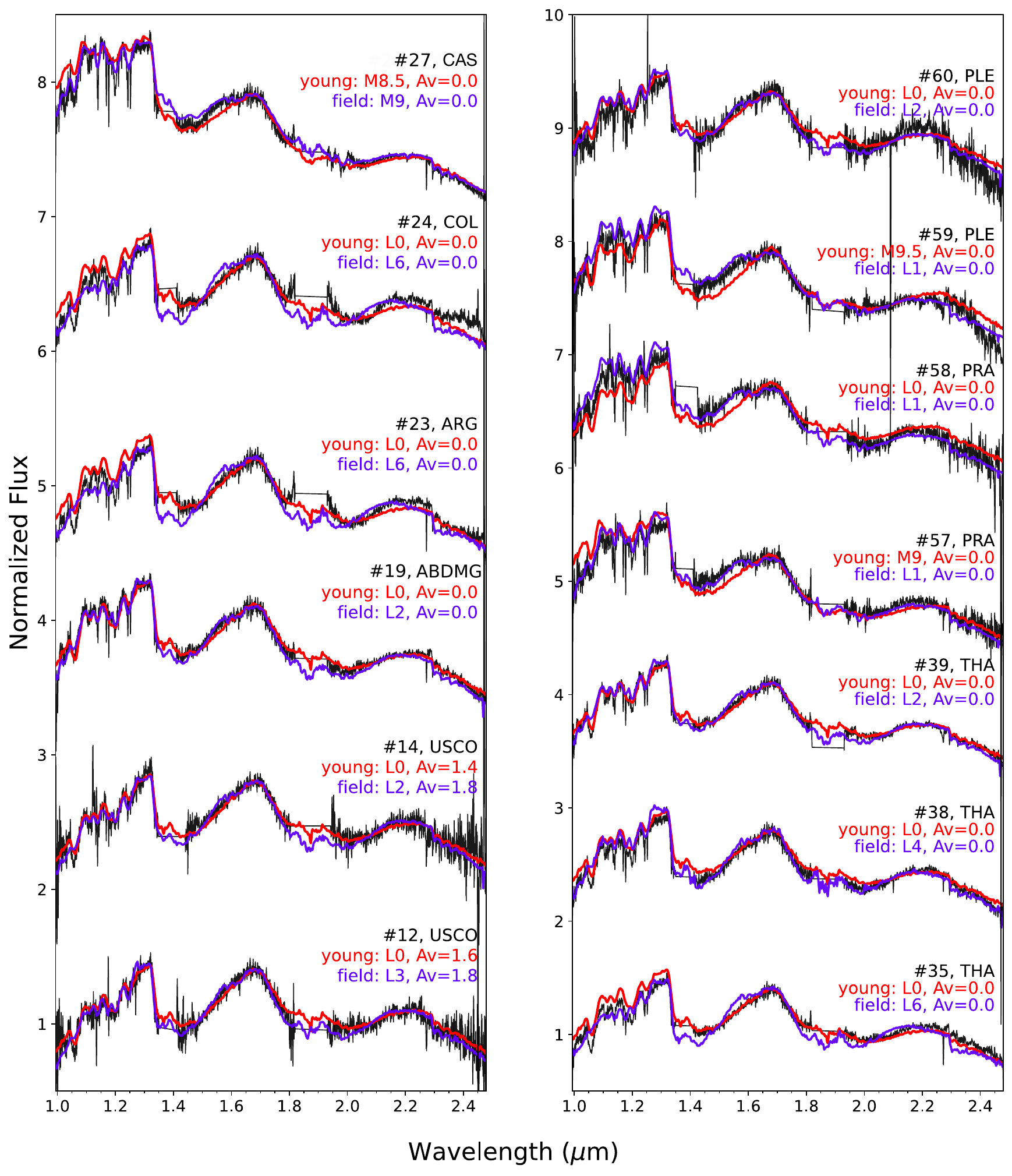}
\caption{Comparison between the best-fitted young (red) and field (purple) templates for dwarfs where the spectral classification by direct comparison with templates yields a result of M or L field. The ID and young association are shown above the correspondent spectrum (in black). All spectra are normalized at 1.67 $\mu$m. An offset between the spectra was added for clarity.}
\label{fig:field_vs_young}       
\end{figure}

Fig.~\ref{fig:field_vs_young} shows the affected spectra, along with the best young (red) and field templates (purple). A particularly useful feature to observe here is the broad-band shape of the H-band, which has a well-defined triangular shape in the young templates. Objects \#57 and \#58, located in PRA which is by far the oldest region considered here, are better represented by the field templates. Furthermore, field templates also appear to work well for the somewhat younger sources, such as \#27 located in CAS, and \#59 and \#60 in PLE. For these objects, we retain the SpT attributed by the fitting process. The remaining spectra visually appear to be better represented by the young templates, and therefore we take the best-fit SpT derived using the young template as the final one. 

\begin{figure}
\centering
    \includegraphics[width=0.44\textwidth]{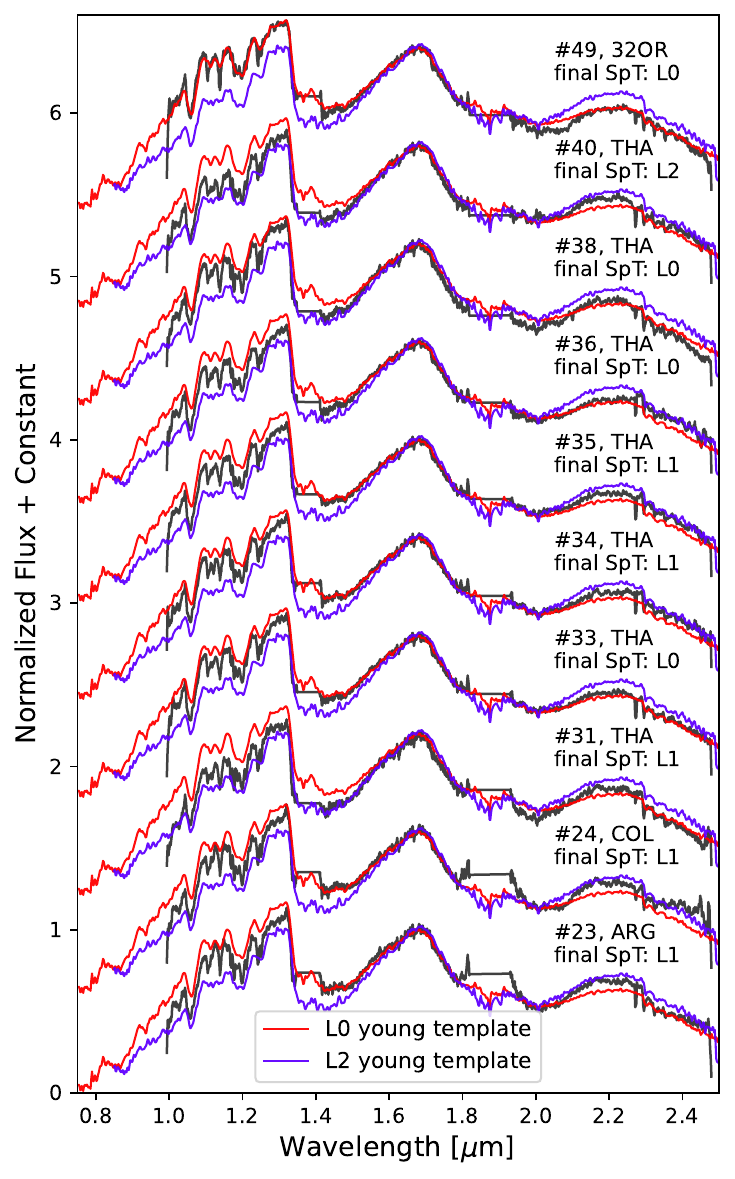}
\caption{Spectra of objects in the SpT range L0-L2 (in black) compared with L0 (red) and L2 (purple) young templates. The spectra were normalized at 1.67 $\mu$m. The ID of each object is listed on the left of the respective spectrum. An offset between the spectra was added for clarity.}
\label{fig:probable_L1}       
\end{figure}

\begin{figure}
\centering
    \includegraphics[width=0.49\textwidth]{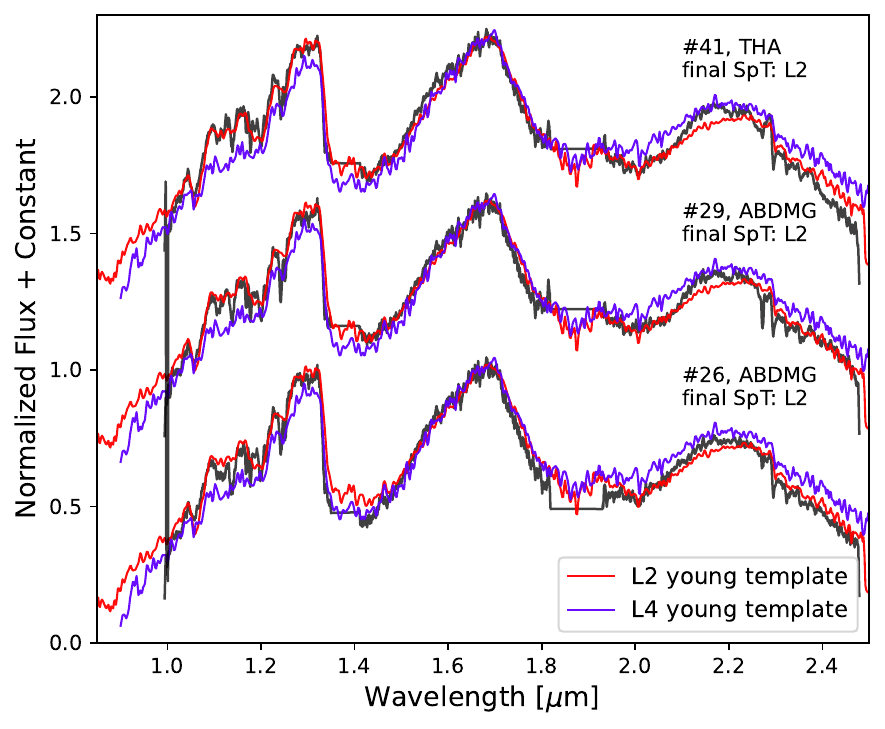}
\caption{Spectra of objects in the SpT range L2-L4 (in black) compared with L2 (red) and L4 (purple) young templates. The spectra were normalized at 1.67 $\mu$m. The ID of each object is listed on the left of the respective spectrum. An offset between the spectra was added for clarity.}
\label{fig:probable_L3}       
\end{figure}

\begin{figure}
\centering
    \includegraphics[width=0.45\textwidth]{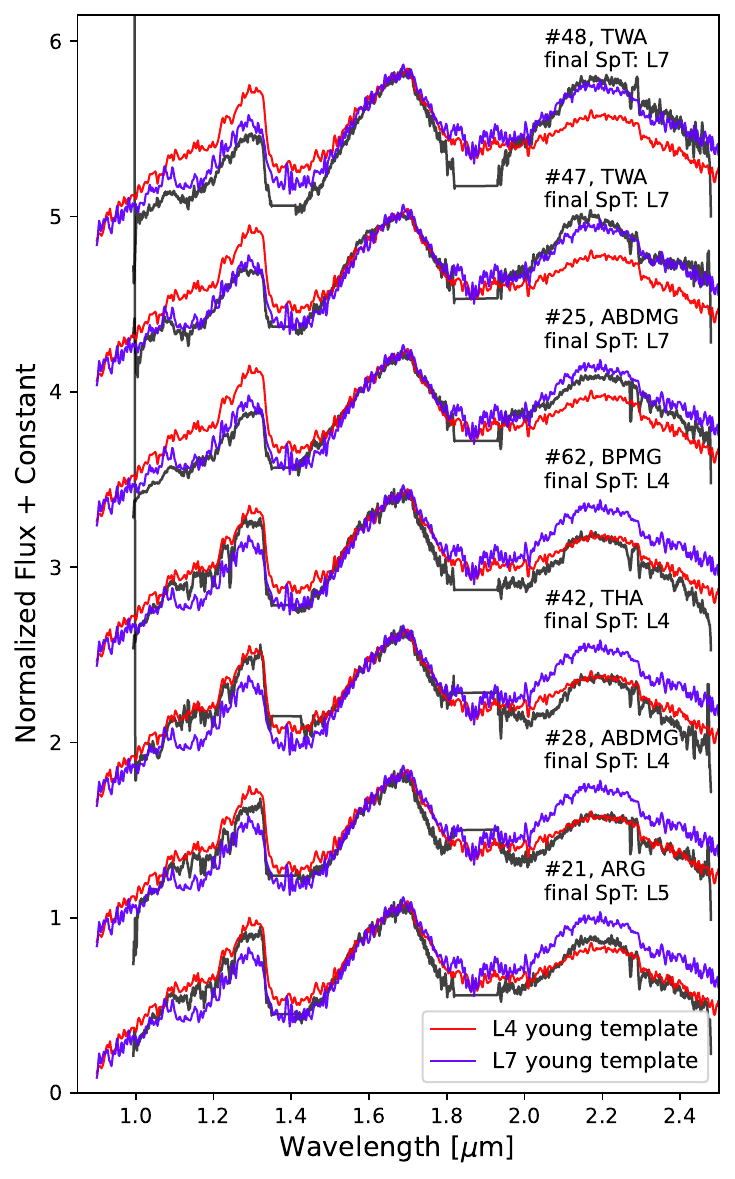}
\caption{Spectra of objects in the SpT range L4-L7 (in black) compared with L4 (red) and L7 (purple) young templates. The spectra were normalized at 1.67 $\mu$m. The ID of each object is listed on the left of the respective spectrum. An offset between the spectra was added for clarity.} 
\label{fig:probable_L5L6}       
\end{figure}

\begin{figure}
\centering
    \includegraphics[width=0.49\textwidth]{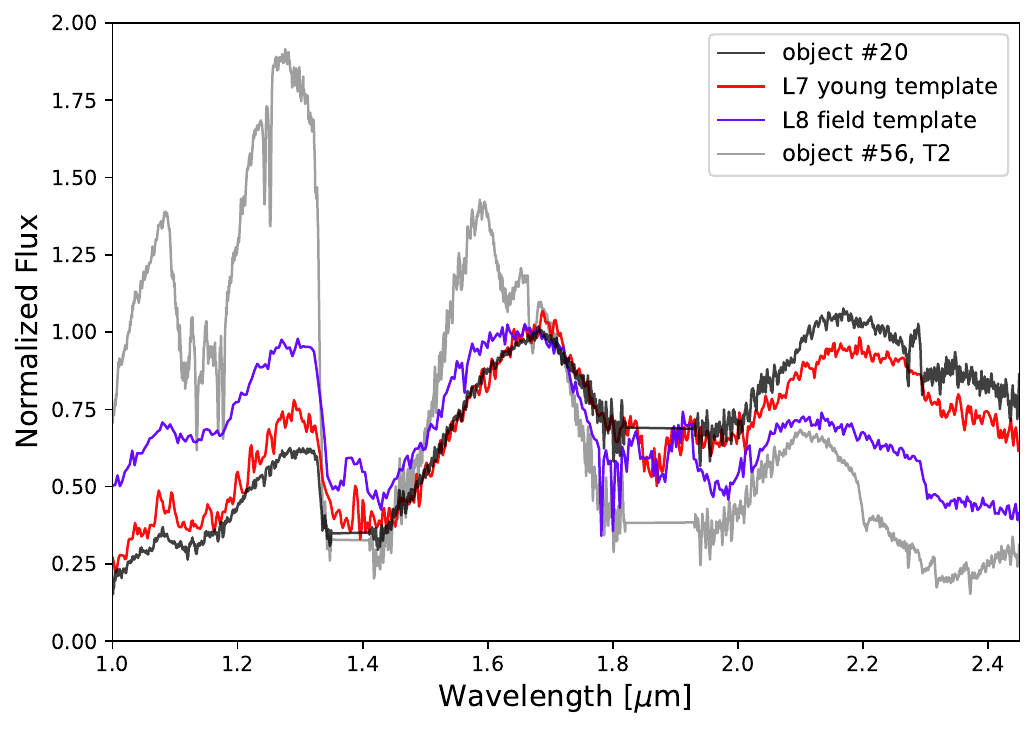}
\caption{Spectrum of the object \#20 (PSO J318.5-22, in black) along with the L7 young template (in red), L8 field template (in purple), and the spectrum of the object \#56 (SIMP J013656.5+093347.3, in gray) which was classified as T2 by direct comparison with spectral templates. The latter is shown to underline the significant change in the spectral morphology associated with the L/T transition. All spectra are normalized at 1.67 $\mu$m.}
\label{fig:object_20}       
\end{figure}

The reason why these spectra originally appeared to be better fitted by the field templates is probably the sparsity of the young template grid. For the same reason, we decide to visually inspect the L-type objects to determine if some of them have a SpT that would appear to be in between the SpTs defined by the grid.
In Fig.~\ref{fig:probable_L1}, we show a set of objects with SpTs in the range L0-L2, and in Fig.~\ref{fig:probable_L3} those with SpTs potentially in the range L2-L4. From Fig.~\ref{fig:probable_L1}, we see that several spectra appear to be in between the L0 and L2 templates, and we, therefore, re-assign them a SpT L1$\pm 1$ (objects \#23, \#24, \#31, \#34 and \#35; see Table~\ref{tab:spec_class}). The range of L2-L4 (Fig.~\ref{fig:probable_L3}) is more difficult to judge visually. 
In most cases, at least two bands are very well fitted by one of the two templates, while the third one (most often the $K-$band) appears a bit off. However, we can neither say that something in between would necessarily provide a better fit. For these objects, we, therefore, retain the young SpT as returned by the fitting procedure.

In Fig.~\ref{fig:probable_L5L6}, we show the objects best fitted by the young L4 or L7 templates. Four of the shown objects (\#21, \#28, \#42 and \#62) are closer to the L4 templates, although most of them do not consistently follow the template shape in all bands. In the case of object \#21, its shape appears to be in between L4 and L7, but closer to L4. We, therefore, assign it a SpT of L5. Four objects were classified as L7 young by direct comparison with spectral templates. The spectra of objects \#25 (2MASS J03552337+1133437), \#47 (2MASS J11193254-1137466) and \#48 (2MASS J114724.10-204021.3) seem well fitted by a L7 young template, although in the case of \#25 both J- and K-bands appear somewhat fainter relative to the H-band when compared to the template. As for object \#20 (PSO J318.5-22), Fig.~\ref{fig:object_20} compares its spectrum (in black) with that of object \#56 (SIMP J013656.5+093347.3, in gray) which is classified as T2, and that of an L7 young (in red) and an L8 field templates (in purple). 
Compared to the young L7 template, object \#20 appears to be of a slightly later SpT, however, it is difficult to assign an exact SpT to it, given the lack of appropriate comparison templates. The L-type field templates are too blue compared to young L-types and thus completely inadequate for spectral typing in this age regime. We show object \#56 from our sample, which is a $\sim200$\,Myr old T2 dwarf, to demonstrate the rapid change in the spectral form at the LT transition. We assign this object a SpT L8$\pm$1, given that it appears later than L7, but is also clearly not a T-dwarf.

Young, low-gravity mid-L and late-L dwarfs typically show
redder near-infrared colors than their older,
higher-gravity counterparts with the same spectral type \citep{Faherty_2016,Liu_2016}. For several T-dwarfs in their sample, \citet{Zhang_2021} report $J-K$ colors that are 0.4 - 0.8 mag higher than what would be expected for field dwarfs. We observe a similar trend for 4 out of 5 T-dwarfs in our sample. Several of these objects also appear as potential binary candidates (see Sect.~\ref{subsec:binaries}). As we will discuss in that section, it seems that the low-gravity behaviour in T-dwarfs may be mimicking binarity when assessed using the field T-dwarf templates.

Finally, we do not perform a similar visual inspection for the L-type objects located in USCO since they may have some degree of extinction, which makes it more difficult to judge visually if some intermediate SpT would be more appropriate (SpT$+$A$_{V}$ degeneracy). The final spectral classifications obtained by comparison with spectral templates and posterior visual inspection, are summarized in Table \ref{tab:spec_class}, column "SpT Templates".

\begin{figure}
\centering
    \includegraphics[width=0.45\textwidth]{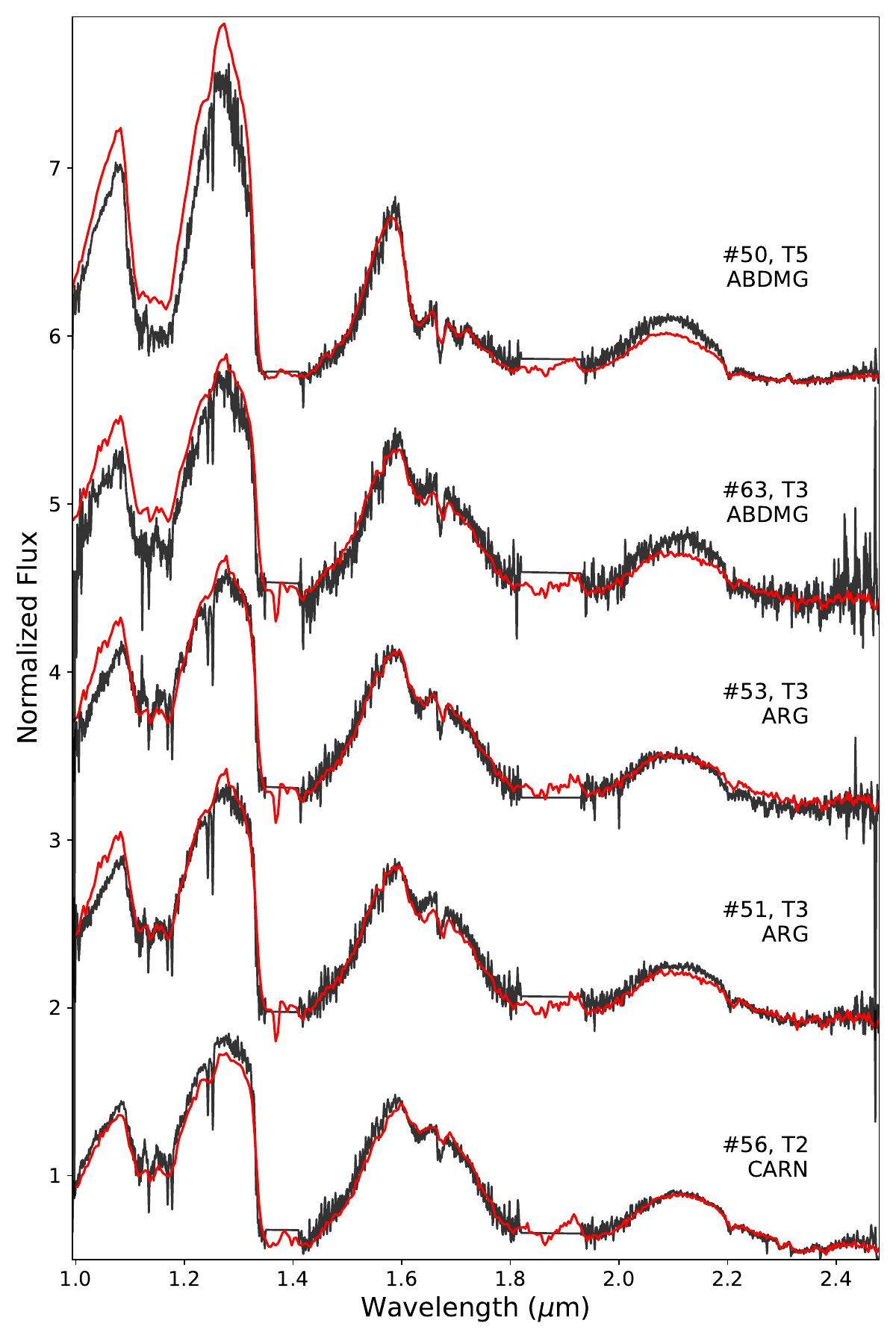}
\caption{Spectra of objects classified as T-type dwarfs along with their best-fitted template (in red). The ID, derived SpT and region of each object are shown next to the correspondent spectrum. All the spectra are normalized at 1.57 $\mu$m. An offset between the spectra was added for clarity.}
\label{fig:Ttype_temp}       
\end{figure}

\begin{figure}
\centering
    \includegraphics[width=0.43\textwidth]{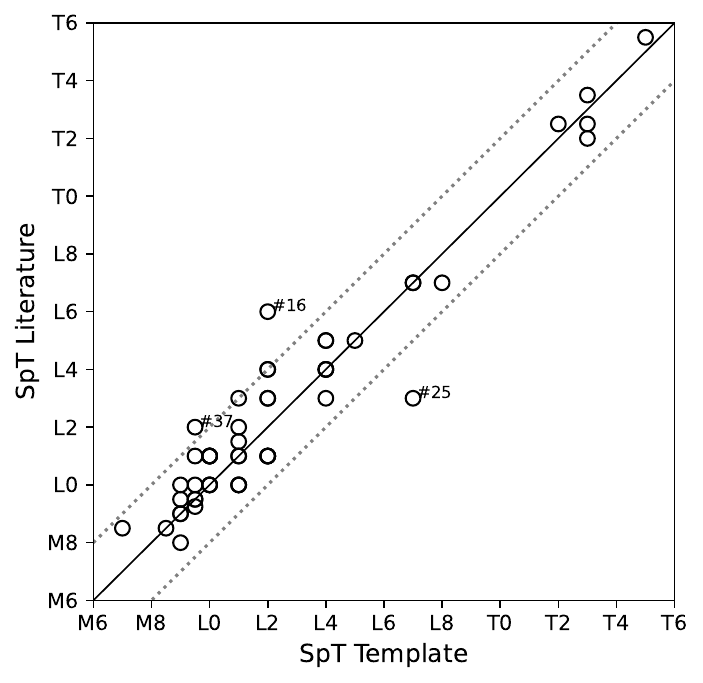}
\caption{Comparison between the derived NIR SpT by direct comparison with spectral templates (see Sect.~\ref{subsec:spt_temp}) and the literature NIR SpT for the X-shooter dataset. The solid black line indicates perfect agreement between classifications. The black dotted lines represent the $\pm$2 sub-SpT range. The ID of the objects for which their classifications do not agree within $\pm$2 sub-SpTs is shown.}
\label{fig:temp_vs_lit}       
\end{figure}

In Fig.~\ref{fig:temp_vs_lit}, we compare the SpTs derived in this section and those obtained from the literature. The solid lines indicate a perfect agreement, while the dotted lines mark $\pm2$ sub-SpT uncertainty. We see that the majority of our results agree well with the literature classification, except for 3 objects. Object \#16 belongs to USCO and its literature classification comes from \citet{Lodieu_2018}, which does not consider extinction when deriving SpTs. The effect of increasing the extinction in the spectra is similar to the effect of decreasing the effective temperature, i.e. by ignoring the extinction, one will obtain systematically later SpTs\footnote{This in fact is the case for all other USCO objects as well, whose X-shooter spectra have first been analyzed in \citet{Lodieu_2018}. For this reason, as a literature value in Table~\ref{tab:spec_class}, we prefer the classification from \citet{Luhman_2018} which considers extinction in their fitting process.}. As for object \#37 (J0357-4417), it is a known unresolved binary system identified by \citet{Bouy_2003}. \citet{Marocco_2013} estimated the NIR SpTs of its components by fitting synthetic binary templates and the best-retrieved fit was that of an L4.5+L5($\pm$1) composite. On the other hand, in the optical, the SpTs derived for this object were M9+L1 \citep{Martin_2006}. The difference between classifications was assigned to the fact that the NIR composite templates were created from field dwarfs and object \#37 exhibits signs of youth. The assigned unresolved NIR SpT in the literature was L2pec \citep{Marocco_2013}, based on the best fit in the J-band. Our composed SpT derivation is M9.5$\pm$0.5. Nonetheless, we note that the template does not reproduce well the object's spectrum (see Fig. \ref{fig_app:other_spec}). Regarding the one object located below the lower dotted line (object \#25), it was classified as an L3 dwarf in \citet{Allers_Liu_2013} by combining the information from direct comparison with spectral templates and index classifications. However, when analyzing individually the comparison between moderate resolution spectra and templates for the J- and K-bands separately, they obtained the best match of L7 and L2, respectively. In this work, we estimate a spectral classification of L7.0$\pm$1.0 by direct comparison with spectral templates. Nonetheless, we note that although the H-band is well fitted, both the J- and K- bands are redder than the template fitted. An important point to note is that different normalization wavelengths can yield slightly different spectral classifications. In this work, we use the H-band, however, some other works, compare their template fittings by normalizing the spectra in the J-band (e.g, \citealt{Kirkpatrick_2010}).

\subsection{Unresolved binaries}
\label{subsec:binaries}

\begin{figure*}
\centering
    \includegraphics[width=0.85\textwidth]{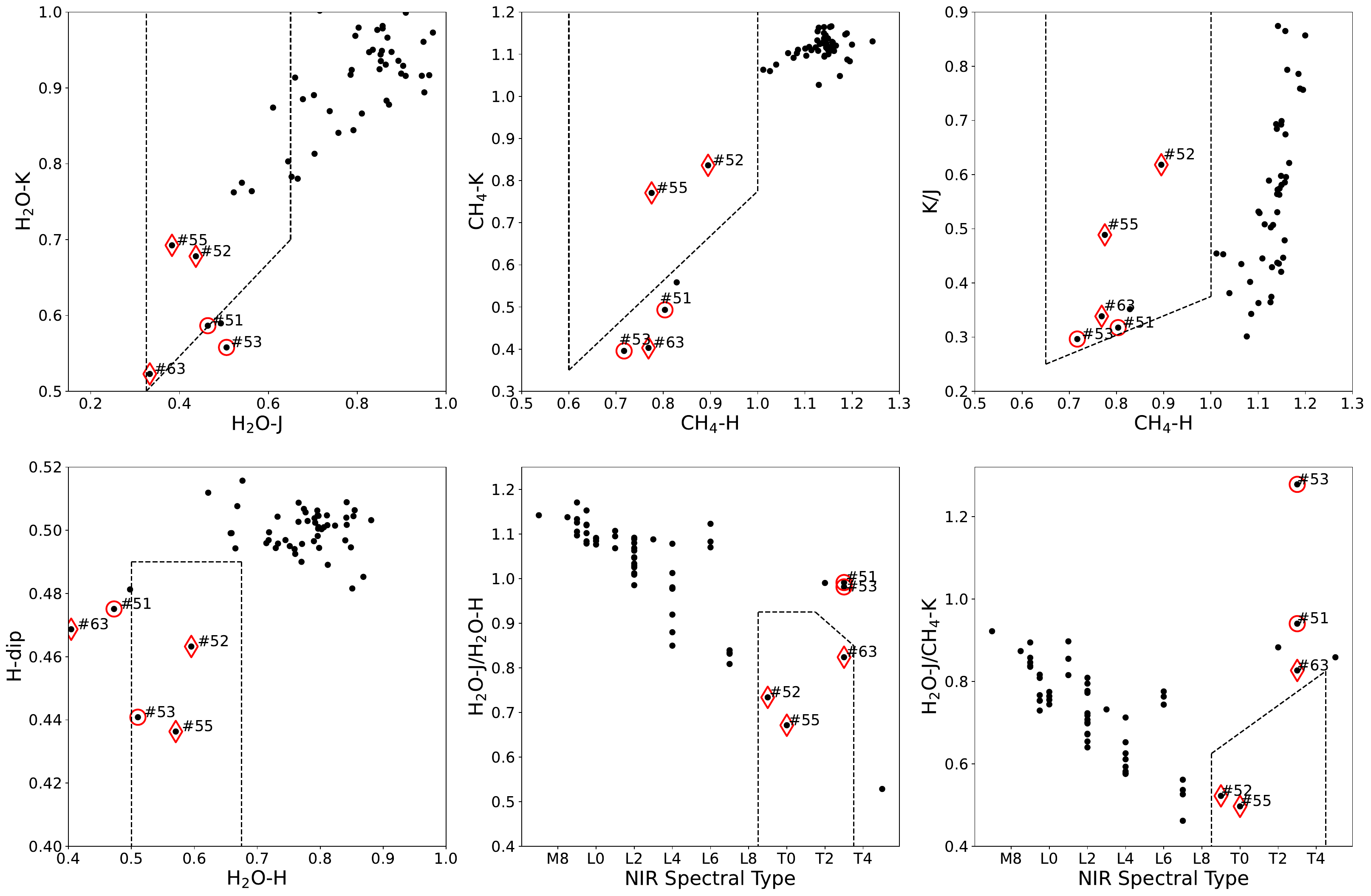}
\caption{Spectral index criteria applied for unresolved binary candidates selection defined in \citet{Burgasser_2010}. The top three panels and the first bottom panel compare pairs of index values whereas the other panels compare index ratios to the NIR SpT derived in Sect. \ref{subsec:spt_temp}. The objects from our dataset are represented as black dots. Strong binary candidates are highlighted with red diamonds and weak binary candidates with red circles.}
\label{fig:binarity}       
\end{figure*}

\begin{figure*}
  \centering
  \includegraphics[width=.27\textwidth]{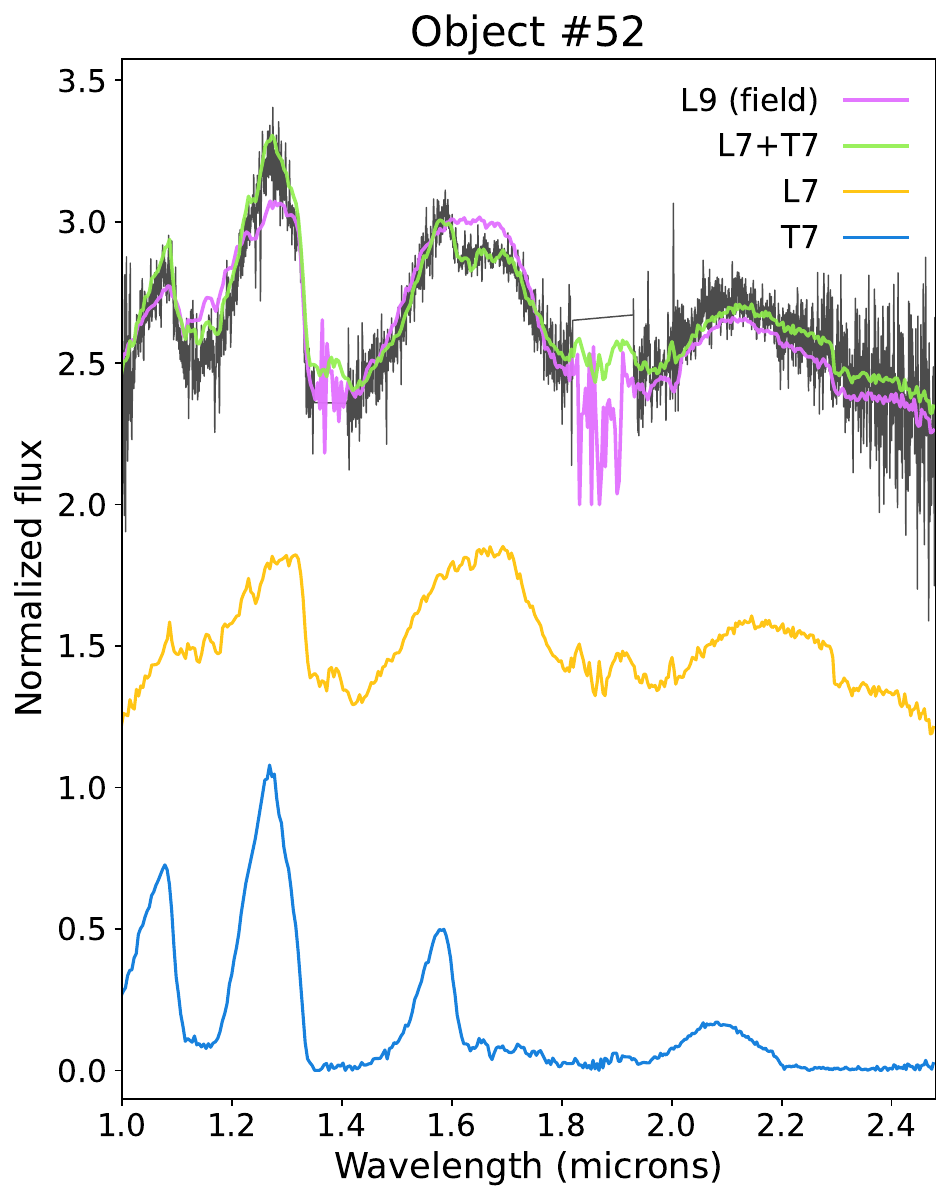}
  \hspace{0.1cm}
  \includegraphics[width=.27\textwidth]{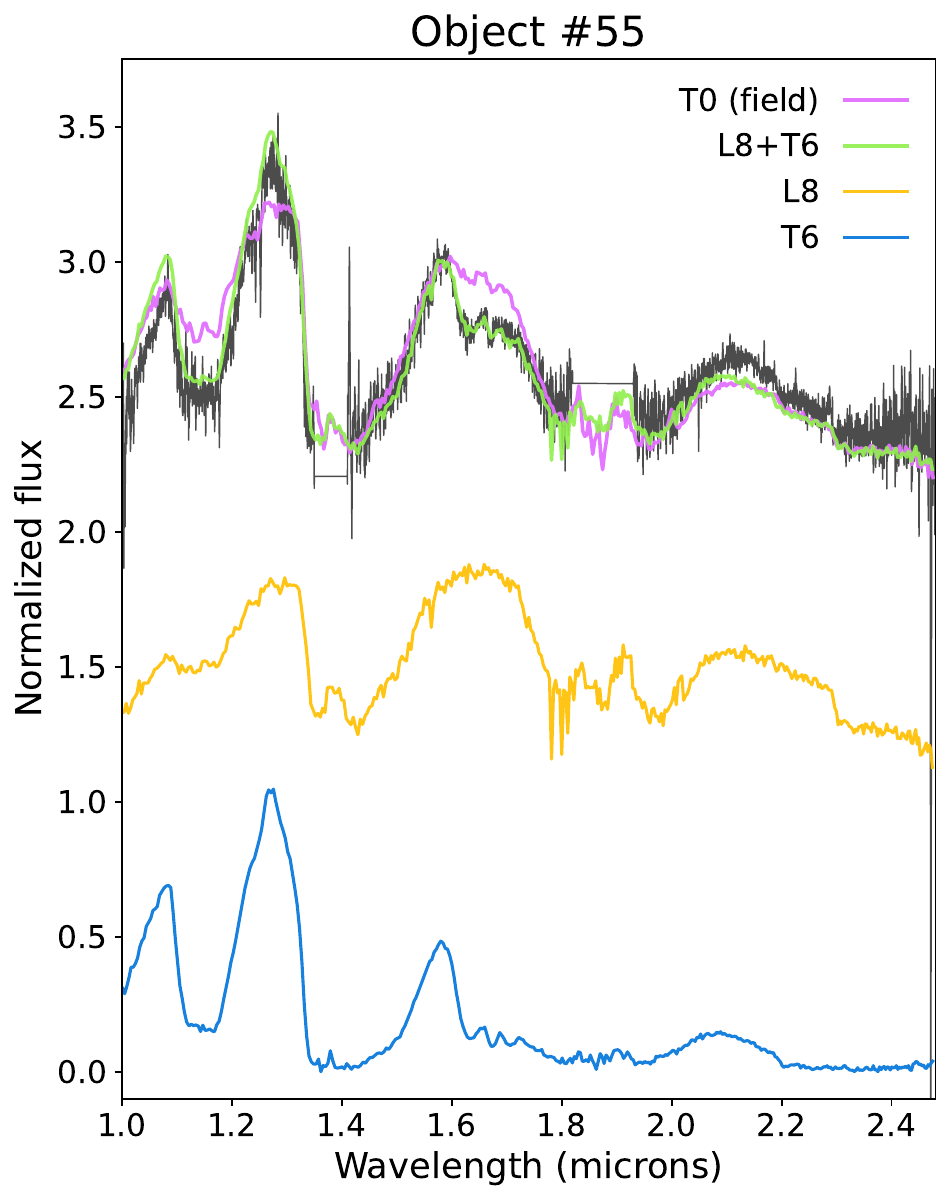}
  \hspace{0.1cm}
  \includegraphics[width=.27\textwidth]{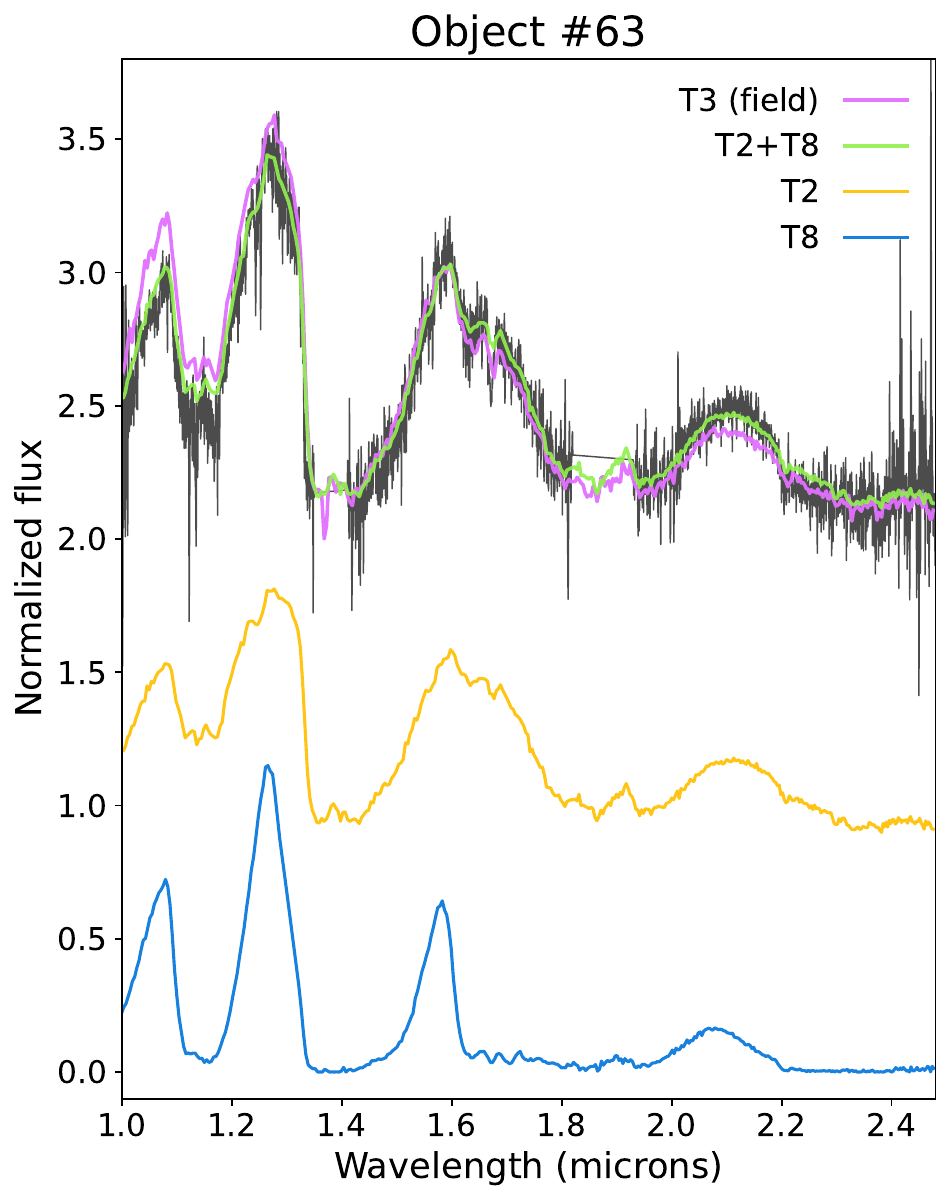}
  \vspace{0.5cm}
  \includegraphics[width=.27\textwidth]{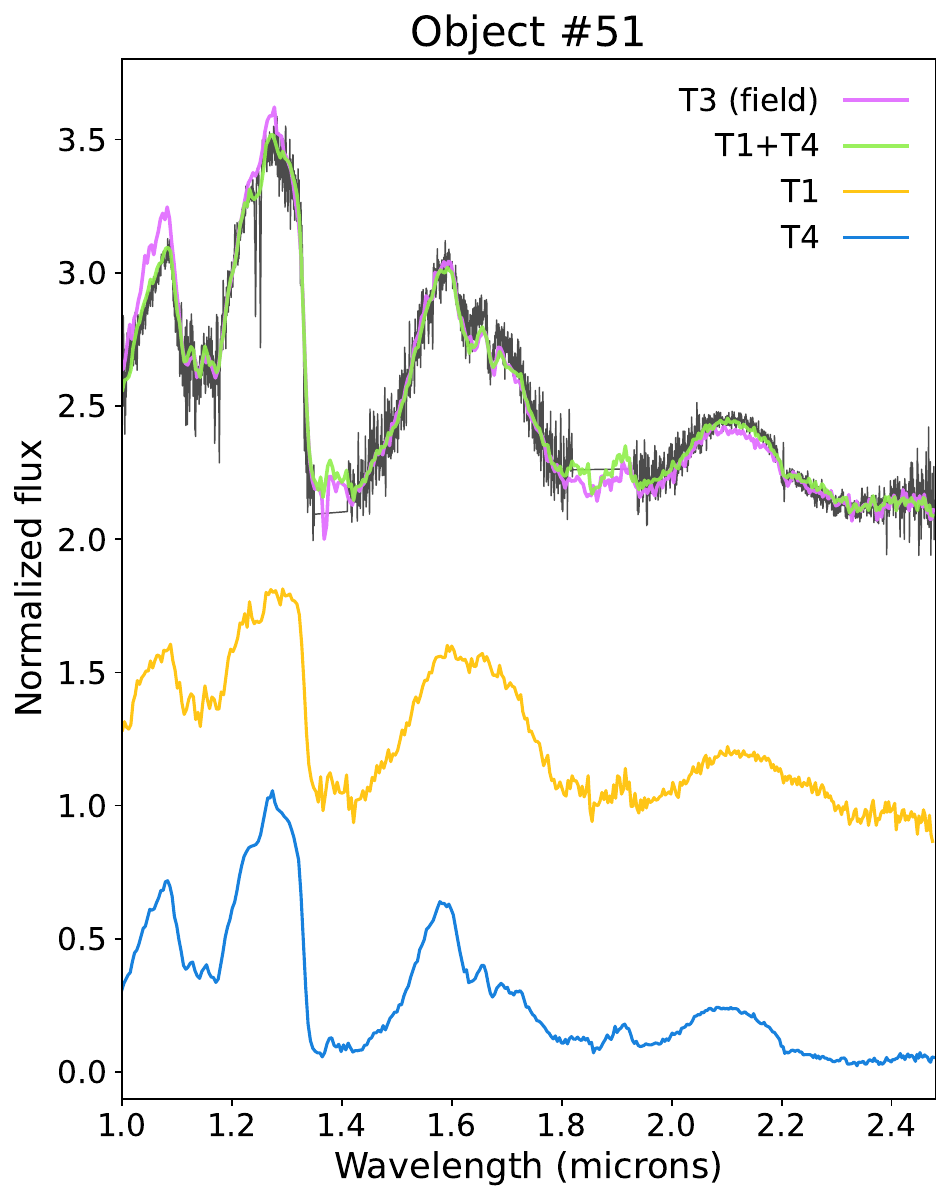}
  \hspace{0.1cm}
  \includegraphics[width=.27\textwidth]{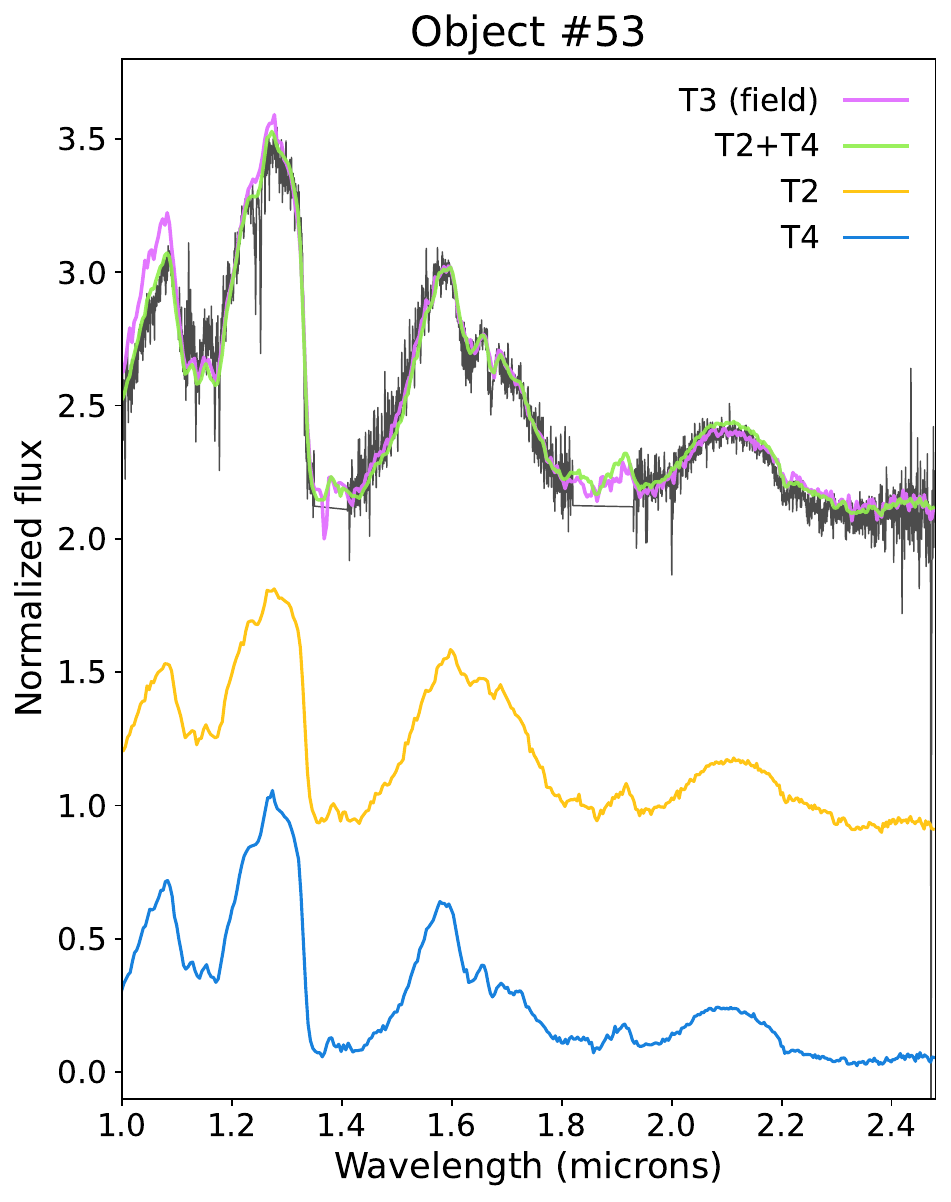}

  \caption{Strong (objects \#52, \#55 and \#63) and weak (objects \#51 and \#53) binary candidates (in black) overplotted with the best-fitted individual templates (magenta) and the best-fitted composite templates (green). The two components of the synthetic binary templates are also plotted in yellow and blue, respectively.}
  \label{fig:composite_temp}
\end{figure*}

Some objects in our sample show peculiar spectra which do not provide a good fit when compared to any of the spectral templates. One of the main reasons that might explain the observed peculiarity of a spectrum is multiplicity. The transition from L- to T-type objects supposes significant changes in spectral properties, caused by pronounced, deep molecular absorption bands (e.g. CH$_4$, H$_2$O) developing at the lower temperatures.   
Hence, a blended-light spectrum from an L/T binary combination will appear peculiar when compared to spectra of other ultracool dwarfs \citep[e.g.][]{Bardalez_2014,Ashraf_2022}.

To test whether our sample harbors potential unresolved binaries, we used the method defined in \citet{Burgasser_2010}, based on six different combinations of index-index and index-SpT diagrams that segregate possible unresolved LT-dwarf binaries. If an object matches two of the six criteria it is selected as a weak binary candidate, whereas satisfying at least three criteria, places the object among the strong candidates. This method is especially sensitive to companions that differ in more than 1 SpT (i.e. a combination of an L-type and a T-type object).

The selection process is shown in Fig.~\ref{fig:binarity}, where six criteria for identifying potential binarity are delineated by dashed lines. We have identified five binary candidates: objects \#52 and \#55 were selected by six indices and object \#63 by three indices therefore these are considered strong candidates; the weak candidates are objects \#51 and \#53 which were selected by two indices each. The strong and weak binary candidates are highlighted as red diamonds and circles, respectively.

To estimate the SpT of both companions of the binary candidates, we use synthetic binary templates, similar to the technique described in \citet{Day-Jones_2013}. For this exercise, we combine the T-type templates with the L-type field templates, since the young L-type templates do not have a complete sequence. The templates were normalized to 1 at 1.28 $\mu$m and scaled to a common flux level using the M$_{J}$-SpT relation from \citet{Faherty_2016}.
Fig.~\ref{fig:composite_temp} exhibits the best-fitted individual templates (in magenta) along with the best-fitted composite templates (in green). The two different components of the synthetic binary templates are shown in yellow and blue. Table \ref{tab:ind_vs_comp} summarizes the SpTs derived by these two different comparisons, complemented with the respective $\chi^{2}$ parameters.

The three strong binary candidates have previously been reported as such. The object \#52 has been disentangled into an L8+T7 binary system by \citet[][BLRT15]{Day-Jones_2013}, in agreement with the SpTs estimated here. \#55 has been reported as an L6+T5 composite in \citet[][BLRT203]{Marocco_2015}, with both SpTs slightly earlier than those found here. It was also selected as a strong binary candidate by \citet{Zhang_2021} with a composite spectral classification of L7+T6. In their discovery paper, \citet{Naud_2014} identified the T-type companion to the young M3 star GU Psc (\#63 here) as a weak binary candidate, following the same set of criteria as we did. Its spectrum appears to be more similar to the confirmed T1+T5 tight binary SDSS J102109.69−030420.1 \citep{Burgasser_2006b} and a T2+T6 binary candidate J121440.95+631643.4 \citep{Geissler_2011} than to any of the single templates.

Similar to \#63, the two weak candidates (\#51 and \#53) appear as composites of two T-type objects\footnote{The disentanglement of these objects' spectra into two T-type objects is in agreement with the combined classifications estimated in \citet{Zhang_2021}: object \#51 was flagged as a weak binary candidate where its combined SpT was estimated as T2+T3, and object \#53 was flagged as a strong candidate with T1.5+T3.5.}. Particularly, in the case of object \#53, appearing to be a combination of two objects with similar SpTs (T2 and T4), the improvement in the fit is only marginal, making its binary status highly questionable. 
We decide to retain those three objects for the remainder of the analysis since they represent one of the very few known T-type objects with low surface gravity. Our understanding of these cool atmospheres at young ages is rather limited, and, since all the claims of the composite nature of the objects come from the field templates, one cannot exclude that their slight peculiarity with respect to the standards may be related to the surface gravity. 
Three out of five T-dwarfs in our sample appear redder than their field counterpart(\#51, \#63 and \#50). Objects \#51 and \#63 were selected as binary candidates, along with object \#53. The remaining object, \#50, has a later SpT (T5), which is not included in the binarity criteria of \citet{Burgasser_2010}.
Interestingly, high-resolution imaging reveals that binary factions among the T-dwarfs are of the order 10\% \citep{Burgasser2003, Burgasser_2006b,Fontanive_2018}, making it highly unlikely that at least 60\% of our sample of T-dwarfs, however small, should indeed be binaries. Considering this, the lower surface gravity of the T dwarfs in our sample might be mimicking binarity.

\begin{table}
\centering
\caption{SpT derived for the strong and weak binary candidates by direct comparison with single and composite templates.}
\resizebox{0.46\textwidth}{!}{
\begin{tabular}{lcccc}
\hline \hline
ID & \begin{tabular}[c]{@{}c@{}}Best fit single \\ template\end{tabular} & \multicolumn{1}{l}{$\chi^{2}_{single}$} & \begin{tabular}[c]{@{}c@{}}Best fit composite\\ template\end{tabular} & \multicolumn{1}{l}{$\chi^{2}_{comp}$} \\ \hline
\multicolumn{5}{c}{Strong Binary Candidates} \\
\#52 & L9 ($\pm$1) & 0.3285 & L7+T7 ($\pm$1) & 0.0696 \\
\#55 & T0 ($\pm$1) & 2.1397 & L8+T6 ($\pm$1) & 0.5694 \\
\#63 & T3 ($\pm$1) & 0.0786 & T2+T8 ($\pm$1) & 0.0327 \\
\hline
\multicolumn{5}{c}{Weak Binary Candidates} \\
\#51 & T3 ($\pm$1) & 1.4887 & T1+T4 ($\pm$1) & 0.4534 \\
\#53 & T3 ($\pm$1) & 0.4758 & T2+T4 ($\pm$1) & 0.4269 \\
 \hline
\end{tabular}}
\label{tab:ind_vs_comp}
\end{table}

This selection method was defined to select binary candidates that straddle the L/T transition. Consequently, there are other objects in our sample that can be potential binary candidates but are not selected as such as a result of their similar SpTs. As seen previously (see discussion in Sect.~\ref{subsec:spt_temp}), object \#37 is an example of such. Another case is object \#47 (J1119-1137), a tight binary where the components have approximately equal magnitudes resulting in an unresolved SpT of L7 \citep{Best_2017}. As the spectral composition of objects with L classifications does not result in such spectral morphological changes as an L+T dwarfs composition, we do not perform a methodical search for these. Object \#37 and \#47 are therefore also retained in the following analysis such as the weak selected binary candidates.

\subsection{Spectral type indices}\label{subsec:spt_indices}

\begin{figure*}
\centering
    \includegraphics[width=\textwidth]{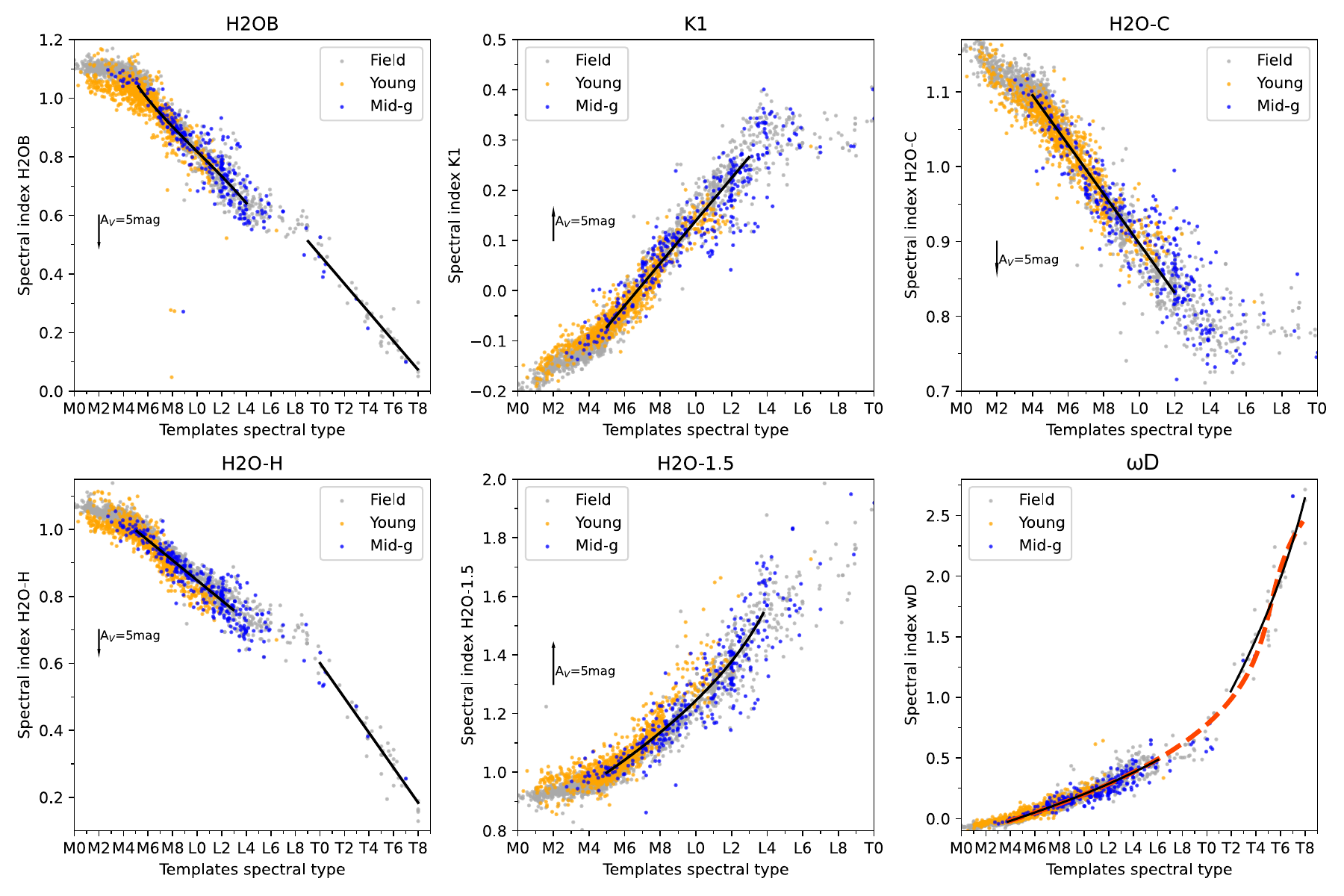}
\caption{Selected SpT indices from the literature versus the NIR SpT derived by direct comparison with spectral templates. The classes young, mid-gravity and field from the dataset of AA22 are represented as orange, blue and gray circles, respectively. The black solid lines represent the sensitivity range of each index (see Table \ref{tab:spt_indices}). The red dashed line in the spectral index $\omega_{D}$ represents the best fit if the sensitivity range is considered from M4 down to T8. The black arrows demonstrate the effect of 5 magnitudes of extinction on the index values. }
\label{fig:spt_indices_ML}       
\end{figure*}

\begin{figure*}
\centering
    \includegraphics[width=0.95\textwidth]{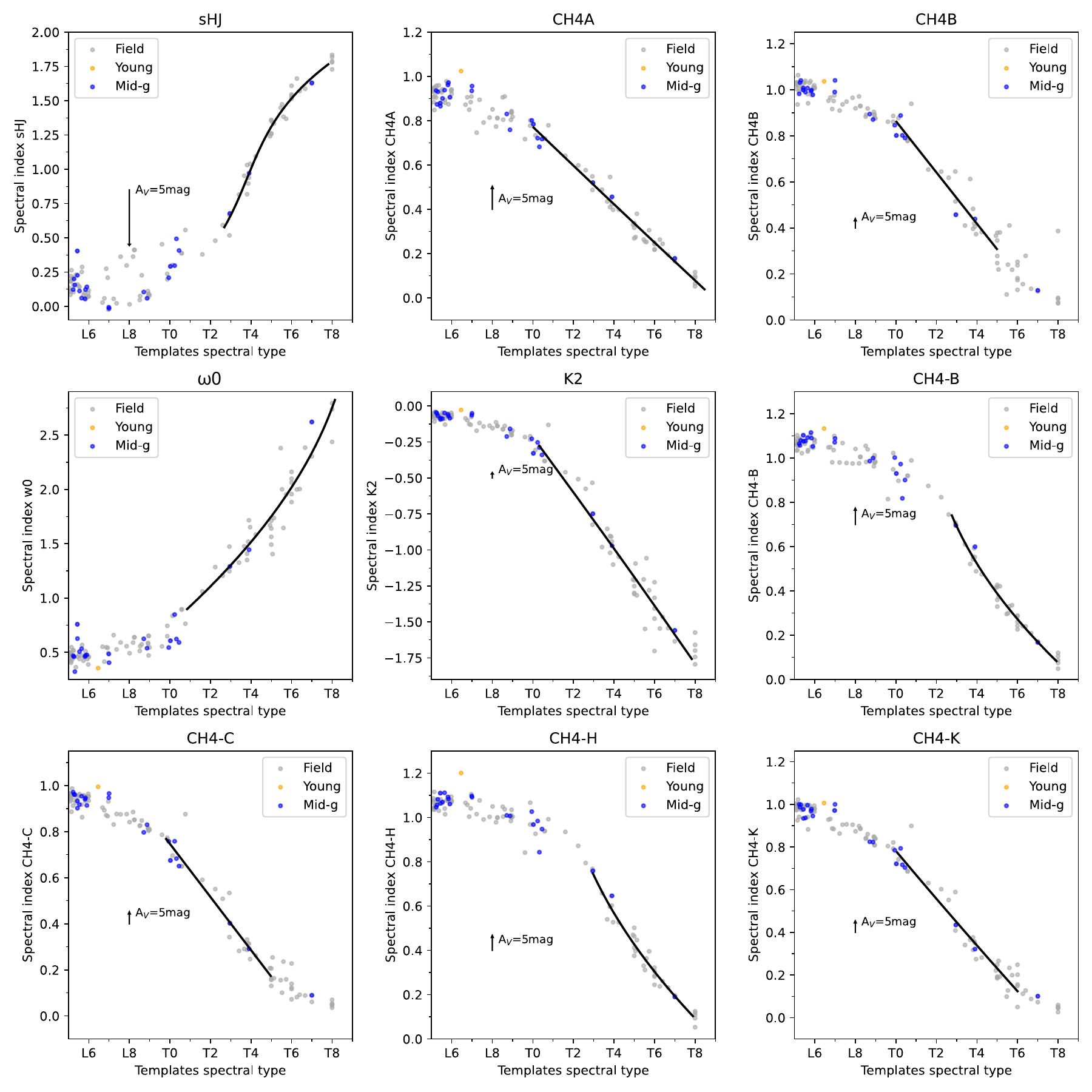}
\caption{Selected SpT indices from the literature versus the NIR SpT derived by direct comparison with spectral templates. The classes young, mid-gravity and field from the dataset of AA22 are represented as orange, blue and gray circles, respectively. The black solid lines represent the sensitivity range of each index (see Table \ref{tab:spt_indices}). The black arrows demonstrate the effect of 5 magnitudes of extinction on the index values.}
\label{fig:spt_indices_LT}       
\end{figure*}

An alternative way to derive SpT is by using spectral indices that correlate well within a given SpT range. This approach is of great interest when, for example, there is no access to the entire NIR wavelength range of the object's spectrum, leading to a decrease in the accuracy of the SpT derivation by direct comparison with spectral templates. Moreover, quantifying specific spectral features does not require an age estimate (if defined as gravity-insensitive) and is less computationally costly. Therefore, it is of utmost relevance to test several indices in order to inspect which are their sensitivity ranges and whether they can be used for spectral classification. One thing to consider is that the main disadvantage regarding the use of spectral indices is the fact that the majority of them are extinction-dependent. Hence, extinction needs to be estimated using another procedure. 

In total 15 indices matched the criteria mentioned above: H2OB, K1, H2O-C, H2O-H, H2O-1.5, $\omega_{0}$, $\omega_{D}$, sHJ, CH4A, CH4B, K2, CH4-B, CH4-C, CH4-H and CH4-K. Figs.~\ref{fig:spt_indices_ML} and \ref{fig:spt_indices_LT} show the indices selected along with their sensitivity range (solid black line) and the effect that an additional extinction of five magnitudes would have on the values of the indices (black arrow). Table \ref{tab:spt_indices} summarizes the ranges of applicability, the type of polynomial fit along with the respective coefficients, and the rmse of each selected SpT index.

\begin{table*}
\centering
    \caption{Spectral type indices selected.}
    \resizebox{0.87\textwidth}{!}{
        \begin{tabular}{l c c c c c c c c}
        \hline \hline
        Index & Index & Sensitivity & \multicolumn{4}{c}{Coefficients$^{a}$} & Type & rmse \\ \cline{4-7}
         name & reference & range & c$_0$ & c$_1$ & c$_2$ & c$_3$ & of fit & (SpT) \\
        \hline
           H2OB & (1) & M5-L4 & 5.808 & 64.156 & -108.775 & 44.736 & 3$^{rd}$ degree & 0.99 \\
             & & L9-T8 & 29.479 & -20.416 & -- & -- & linear & 1.00 \\
            K1 & (2) & M5-L3 & 6.707 &  23.623 & -- & -- & linear & 0.77 \\
            H2O-C & (3) & M4-L2 & 37.130 & -30.240 & -- & -- & linear & 0.79 \\
           H2O-H & (4) & M5-L3 & 38.348 & -33.438 & -- & -- & linear & 0.98 \\
            &  & T0-T8 & 31.521 & -19.174 & -- & -- & linear & 0.68 \\
           H2O (1.5 $\mu$m) & (5) & M5-L4 & -32.696 & 51.745 & -14.011 & -- & 2$^{nd}$ degree & 0.97 \\
           $\omega_{D}$ & (6) & M4-L6 & 4.547 & 29.312 & -11.631 & -- & 2$^{nd}$ degree & 0.94 \\
           & & T1-T8 & 15.649 & 6.877 & -0.837 & -- & 2$^{nd}$ degree & 0.66 \\

           sHJ & (7) & T2.5-T8 & 18.803 & 10.554 & -8.353 & 2.982 & 3$^{rd}$ degree & 0.25 \\
           CH4A & (1) & T0-T8 & 28.918 & -11.717 & 0.220 & -- & 2$^{nd}$ degree & 0.37 \\
           CH4B & (1) & T0-T5 & 27.750 & -8.969 & -- & -- & linear & 0.56 \\
           $\omega_{0}$ & (6) & T1-T8 & 14.897 & 7.534 & -1.007 & -- & 2$^{nd}$ degree & 0.60 \\
           K2 & (2) & T0-T8 & 18.861 & -5.297 & -0.108 & -- & 2$^{nd}$ degree & 0.56 \\
           CH4-B & (3) & T3-T8 & 28.871 & -11.882 & 4.908 & -- & 2$^{nd}$ degree & 0.23 \\
           CH4-C & (3) & T0-T5 & 26.477 & -8.668 & -- & -- & linear & 0.52 \\
           CH4-H & (4) & T3-T8 & 28.985 & -11.037 & 3.981 & -- & 2$^{nd}$ degree & 0.27 \\
           CH4-K & (4) & T0-T6 & 27.176 & -9.550 & 0.473 & -- & 2$^{nd}$ degree & 0.52 \\
        \hline
        \end{tabular}}
\tablefoot{References: (1) \citet{McLean_2003}, (2) \citet{Tokunaga_1999}, (3) \citet{Burgasser_2002}, (4) \citet{Burgasser_2006}, (5) \citet{Geballe_2002}, (6) \citet{Zhang_2018}, (7) \citet{Testi_2001}.

$^{a}$ The spectral type is calculated from the polynomial fits as: SpT = $\sum_{i=0}^{n} c_{i} \times$ (index)$^{i}$ (with $n=1$ for linear, $n=2$ for the second-degree and $n=3$ for the third-degree polynomial fit).}
\label{tab:spt_indices}
\end{table*}

We then proceeded to the computation of the selected indices for each object of our sample. For the objects in USCO, we considered the extinction value estimated in Sect.~\ref{subsec:spt_temp} to correct the spectra for reddening, using the Cardelli extinction law with $R_{V}$=3.1.
The SpT and its uncertainty were calculated as the weighted mean and weighted standard deviation of the values obtained from various indices, which depend on each index's SpT range of applicability. The final values have been rounded to the next 0.5 sub-SpT, and are presented in Table \ref{tab:spec_class}, column "SpT Indices".

The plots shown in Figs.~\ref{fig:spt_indices_ML} and \ref{fig:spt_indices_LT} show a good correlation with SpTs earlier than $\sim$L6, and later than $\sim$T0. The LT transition range is not well populated and also does not show good behavior in the sense that most indices appear rather flat and/or with high dispersion. The only potentially valuable index for this range may be $\omega_{D}$ (lower right panel of Fig.~\ref{fig:spt_indices_ML}) which has also the advantage of being extinction-independent. For this reason, we try to fit a single polynomial to the entire range of the plot (red dashed line). However, in the range L6-T1 we notice that basically all the points are below the fit line, which means that the index tends to provide systematically earlier SpTs in this spectral range. The SpT range L6-T0 is therefore not included in any sensitivity range of the spectral type indices selected.

\begin{figure}
\centering
    \includegraphics[width=0.43\textwidth]{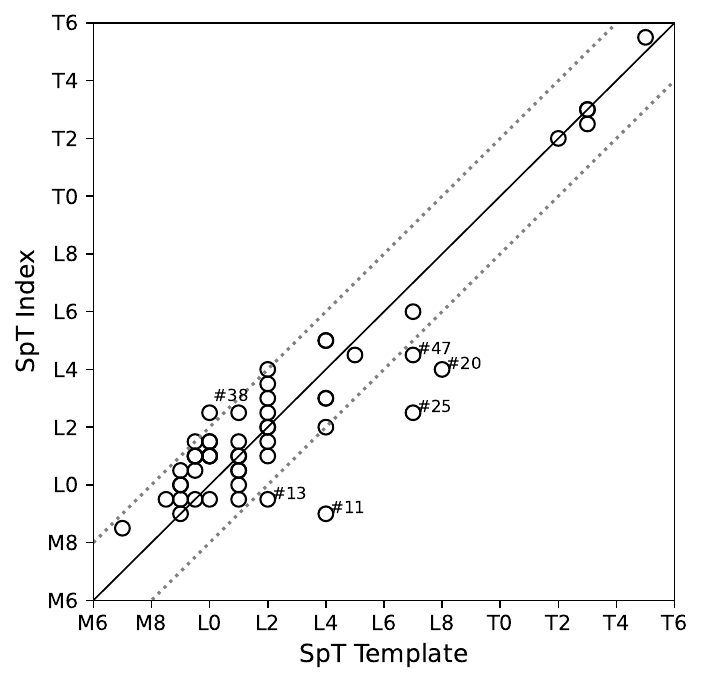}
\caption{Comparison between the derived NIR SpT using spectral indices (see Sect.~\ref{subsec:spt_indices}) and the SpT derived by direct comparison with spectral templates (see Sect.~\ref{subsec:spt_temp}). The solid black line indicates perfect agreement between classifications. The black dotted lines represent the $\pm$2 sub-SpT range. The ID of the objects for which their classifications do not agree within $\pm$2 sub-SpTs is shown.}
\label{fig:temp_vs_ind}       
\end{figure}

In Fig.~\ref{fig:temp_vs_ind} we compare the estimated SpTs obtained with both classification methods: comparison with spectral templates and spectral indices. The classifications seem to agree well within $\pm$2 sub-SpTs, except for 6 objects in our sample. Three of these six objects (IDs \#20, \#25, and \#47) are classified as late-L dwarfs (L8$\pm$1, L7$\pm$3, and L7$\pm$3, respectively) by direct comparison with spectral templates and posterior visual inspection but have an earlier SpT derived with the index classification method (L4$\pm$3, L2.5$\pm$1.0, and L4.5$\pm$3.0, respectively). This is not unexpected, given the behavior of the indices for the late-L types, as discussed above. 
Objects \#11 and \#13 are located in USCO and exhibit discrepant classifications by more than 2 sub-SpTs as well (L4$^{+3}_{-2}$, L2$\pm$2 using templates and M9$\pm$1, M9.5$\pm$1.0 using spectral types indices, respectively).   
On the other hand, the SpT retrieved from the comparison with spectral templates (L0$\pm$2) for object \#38 is earlier than the one computed using the spectral indices (L2.5$\pm$0.5).
The spectrum of object \#11 is quite noisy in several regions important for index calculation, which is probably responsible for the discrepancy. In the case of the objects \#13 and \#38, we notice that the $J-$band part of the spectrum is not that well reproduced by the best-fit spectral template fits.

\section{Evaluation of youth-related features}
\label{sec:youth}

Visual comparison of NIR spectra of young dwarfs to their field counterparts has revealed a wealth of spectral (atomic and molecular) features that differ with age, directly related to differences in surface gravity of these objects \citep[e.g][]{Lucas_2001,Gorlova_2003,Luhman_2017,Allers_Liu_2013}.
AA22 applied machine learning methods to the low-resolution spectra in the range between M0 and L3, and find that the features with the most importance when it comes to separating low-gravity young dwarfs from their field counterparts are the broad-band shape of the H-band, FeH bands, and the lines of the alkali elements (KI and NaI).  
In this section, we evaluate and discuss various spectral indices sensitive to gravity and their applicability to the spectral range $>$L3. We also measure the pseudo-equivalent widths (pEWs) of various prominent alkali lines and study their behavior with spectral type and age. In the following analysis, we assume the spectral type of the sample to be that estimated by comparison with spectral templates (see Sect.~\ref{subsec:spt_temp}).

\subsection{Gravity-sensitive NIR spectral indices}\label{subsec:grav_indices}

\begin{figure*}
\centering
    \includegraphics[width=\textwidth]{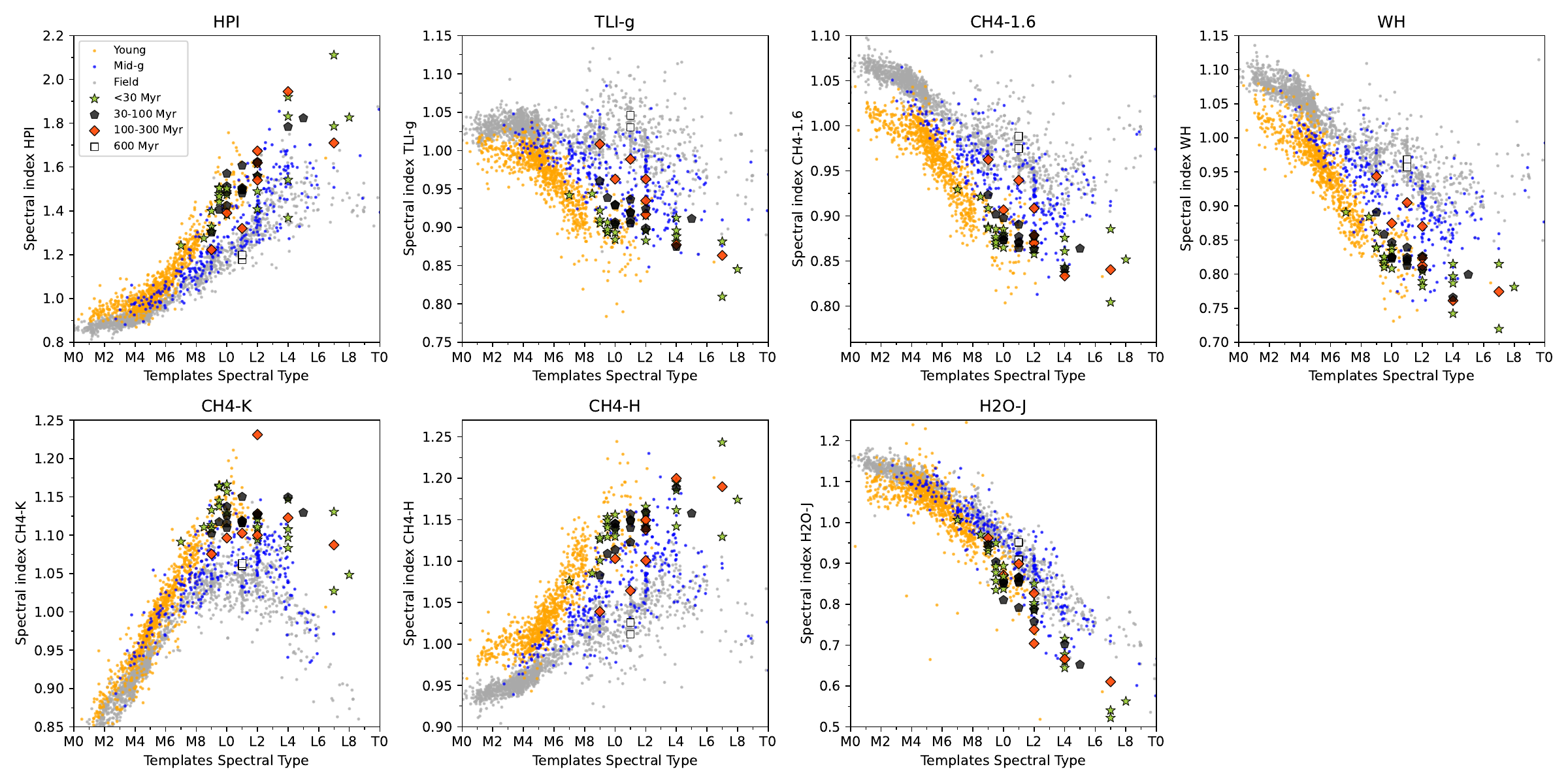}
\caption{Selected gravity-sensitive spectral indices from the literature versus the NIR SpT derived in Sect.~\ref{subsec:spt_temp}. Our sample is divided into the different age classes presented in Sect.~\ref{subsec:sample_selection} where the classes <30\,Myr (USCO, TWA, BPMG and 32OR), 30$-$100\,Myr (ARG, COL and THA), 100$-$300\,Myr (ABDMG, CAS, CARN and PLE), and 600\,Myr (PRA) are shown as green stars, black pentagons, red diamonds, and white squares, respectively. The complementary dataset is divided into young, mid-gravity, and field objects represented as orange, blue, and gray dots, respectively.}
\label{fig:gravsens_indices}       
\end{figure*}

The gravity-sensitive indices that we visually inspected from the literature are listed in Table \ref{tab_app:youth_indices} along with their functional forms, spectral features targeted and references. Some of the indices inspected in Sect.~\ref{subsec:spt_indices} also demonstrated a gravity-sensitive behavior, such as sH2O$^{J}$, CH4 (1.6 $\mu$m), WH, QH, sH2O$^{H1}$, K2, CH4-B, CH4-C, CO, CH4-H and CH4-K. Where applicable, the spectra were corrected for the extinction derived in Sect.~\ref{subsec:spt_temp} using Cardelli's extinction law with R$_{V}$=3.1.
Given the scarcity of low-gravity objects in the late L and T-type regime, this discussion is limited to the spectral types down to L7.

In Fig.~\ref{fig:gravsens_indices}, we show a set of gravity-sensitive indices, using the dataset from AA22, along with the objects analyzed in this work. The orange dots show objects that are typically younger ($<$10\,Myr) than our sample, but we see that they anyway separate well from the field objects. This means that some of these indices will be useful to classify the young members of SFRs in future deep spectroscopic surveys.  

Both HPI and TLI-g are based on the  H$_{2}$O absorption in the H-band.
HPI presents a good correlation with SpT for the younger and field populations from SpT M6, whilst being gravity-sensitive. The fact that a correlation with SpT exists between the different age classes provides an easier way to separate them. 
TLI-g shows a clear separation from, at least, M2 and the intermediate-gravity objects populate the region between the young and field populations. 
Our objects seem to overlap mostly with the intermediate-age class. 
The youngest objects from our sample have at least 10\,Myrs and the young class from AA22 has ages $\lesssim$10\,Myrs. Therefore, the separation might remain if late-L objects from SFRs with ages of only a few Myr follow the trend visible down to L2 for young dwarfs. The objects in the $<$30\,Myr age class later than L4 seem to indicate that this may be so. 
Another water-based index, H2O-J, does not appear gravity-sensitive in the M spectral range but shows a progressively more gravity-sensitive behavior towards the late L-types.

The gravity-sensitive indices that measure the methane features, both in the H- and K- bands, were defined as SpT indices to classify objects in the T spectral sequence. Interestingly, they can have a dual role, because in L-dwarfs the spectral regions considered to compute these indices are actually quantifying the slope of the blue end of each band caused by the water absorption features. The CH$_{4}$-1.6 $\mu$m index considers both absorption peaks in the H-band of T dwarfs enhanced by the methane absorption bands, as does CH$_{4}$-H but with the inverse formulation of CH$_{4}$-1.6 $\mu$m. Both indices increase monotonically through the T sequence and thus can be used to classify T-type objects. However, for SpTs earlier than L6, they demonstrate a good performance as gravity-sensitive indices. The young and field populations can be easily separated since M0, having some overlap within M3 and M5, but for SpTs$>$M6 the separation is very pronounced. WH measures the water absorption feature on the blue side of the H-band and has a similar behavior to CH$_{4}$-1.6 $\mu$m index. CH$_{4}$-K measures the 2.2 $\mu$m methane absorption feature and it does not exhibit a similar behavior as the other selected indices. It does not perform so well in the separation of the different age classes for the early- to mid-M sequence. However, for SpTs later than M8, the trend of the younger and older objects starts to increase and decrease, respectively. This behavior is interesting and could be extremely useful to disentangle young and field populations in the L sequence.

An equivalent analysis of gravity-sensitive indices would be extremely interesting for the T-dwarf range, in light of the upcoming JWST results.
This has, so far, been basically impossible given the lack of known young T-type objects. We note that the T-dwarfs included in this study appear in agreement with the field dwarf positions in plots like that shown in Fig.~\ref{fig:gravsens_indices}.

\subsection{Pseudo-equivalent widths of alkali lines}

\begin{figure*}
\centering
    \includegraphics[width=\textwidth]{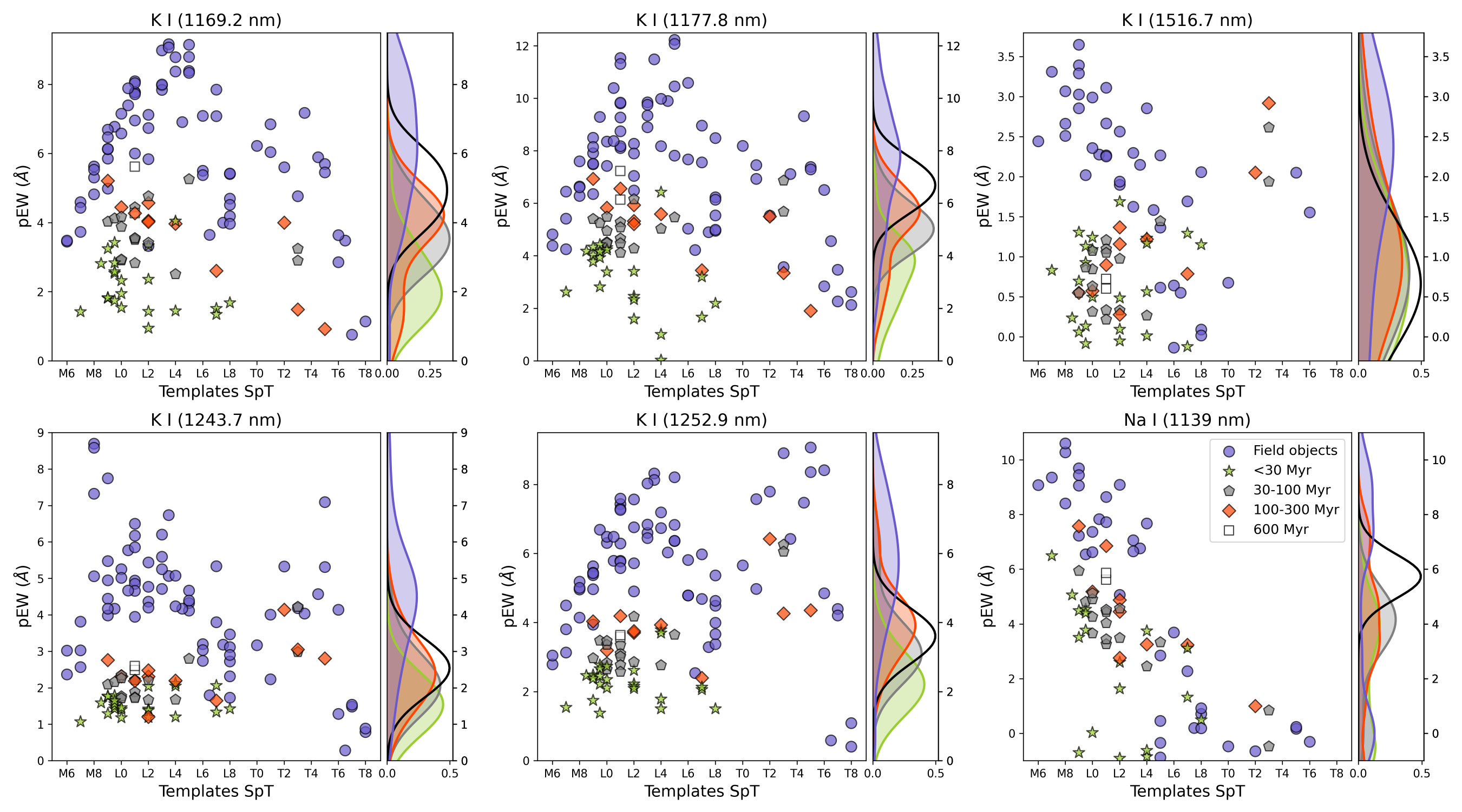}
\caption{Pseudo-equivalent widths of the alkali lines in the J-band as a function of the SpT derived in Sect.~\ref{subsec:spt_temp}. The sample is divided according to the age classes defined in Sect.~\ref{subsec:sample_selection}. The color and marker scheme for our sample is the same as Fig.~\ref{fig:gravsens_indices}. The purple circles are field objects from \citet{McLean_2003} and \citet{Cushing_2005}. In the right panels of each subfigure, there are the KDE for each age class (color-coded accordingly).}
\label{fig:pEW_spt}       
\end{figure*}

\begin{figure*}
\centering
    \includegraphics[width=1\textwidth]{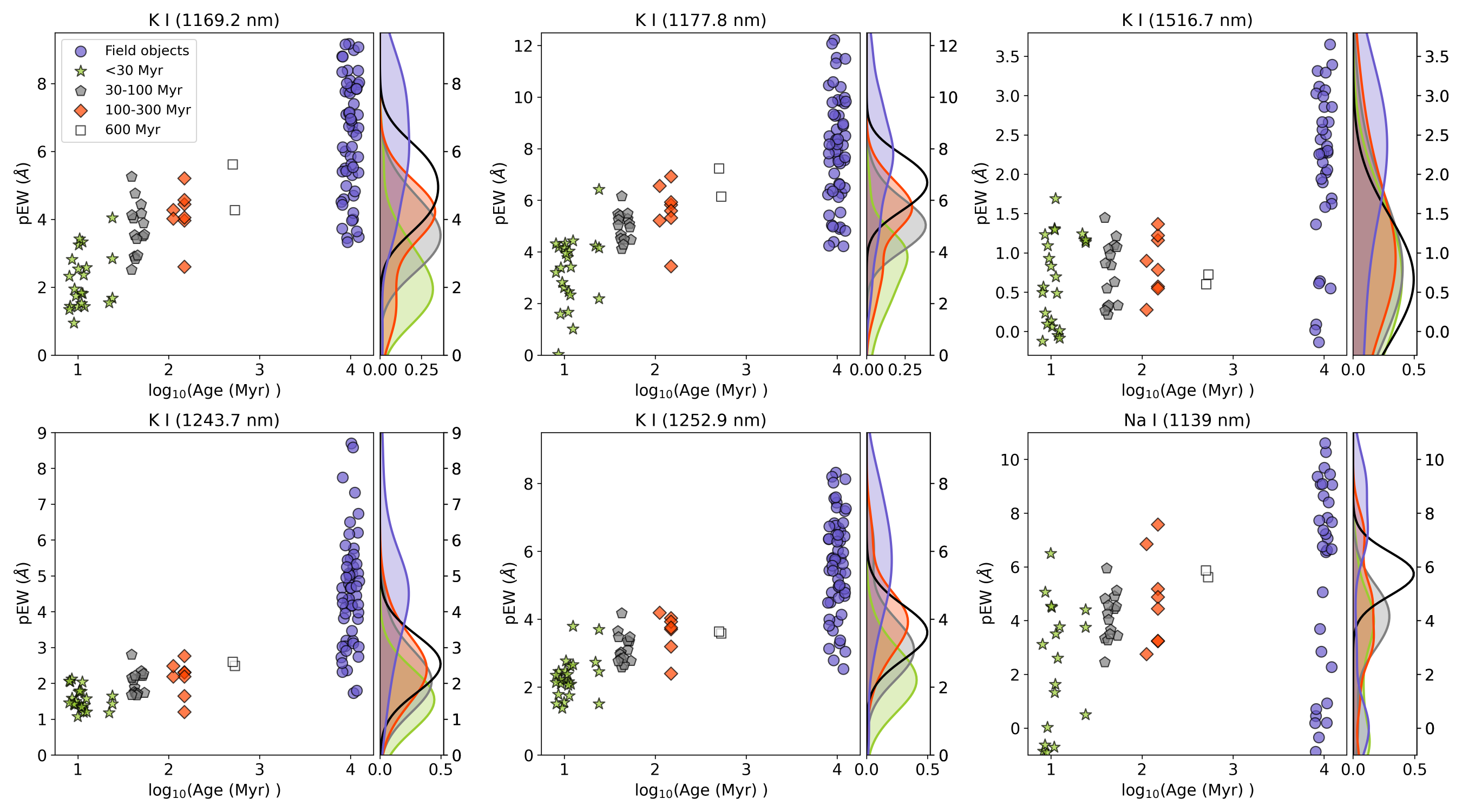}
\caption{Pseudo-equivalent widths of the alkali lines as a function of the age. The sample is divided according to the age classes defined in Sect.~\ref{subsec:sample_selection}. The color and marker scheme for our sample is the same as Fig.~\ref{fig:gravsens_indices}. The purple circles are field objects from \citet{McLean_2003} and \citet{Cushing_2005}. In the right panels of each subfigure, there are the KDE for each age class (color-coded accordingly).}
\label{fig:pEW_age}       
\end{figure*}

The alkali lines (Na\,I and K\,I) are the dominant atomic spectral features in the $J$-band spectra of ultracool dwarfs but are also found in the $H$- and $K$-bands. The strength of the absorption of the alkali lines is primarily affected by the surface temperature of the object, but pressure broadening plays a major role in creating their final shape. They have been found to be a very good probe for differences in surface gravity for late-type objects \citep{Gorlova_2003,Schlieder_2012,Allers_Liu_2013, Martin_2017,Manjavacas_2020}. In this section, we study the pseudo-equivalent widths (pEW) of the alkali lines for the objects in our sample and compare them with that of field objects. We focused on the K\,I doublets at 1.17, 1.24 and 1.51 $\mu$m and the Na\,I doublet at 1.14 $\mu$m. The Na\,I doublet in the K-band (2.21 $\mu$m) basically disappears at L-types \citep{McLean_2003,Cushing_2005}, which is our principal SpT range of interest in this work, and thus was not considered. 

The pEWs were measured using the same method as in \citet{Manara2013}, which integrates the Gaussian fit of the line, previously normalized to the local pseudo-continuum at the edges of
the line. 
The errors in the pEW measurements are estimated from the propagation of 1$\sigma$ of the estimated continuum. We visually checked that the extent of all the lines is correctly estimated, and corrected them if needed. We also inspected the appearance of the alkali lines. Some sources presented contamination from single-pixel spikes within the extent of the line: the pEW of those lines is not considered. We also measured the pEWs of the alkali lines for a sample of field dwarfs with SpTs M6-T8 with a somewhat lower spectral resolution than that of X-shooter (R$\approx$2000), coming from different public surveys \citep{McLean_2003,Cushing_2005}. 

In Fig.~\ref{fig:pEW_spt} we present the pEW of the alkali lines as a function of the SpT. The sources from our X-shooter sample are separated into age classes following the same color scheme used in the rest of this work. Field objects are represented with purple circles. In order to get a clearer view of potential trends in the data, on the right side of each plot we show the distribution for each age class smoothed with a Kernel Density Estimator (KDE). In the L-type range, our objects are clearly distinguished from the field ones, having lower pEW values, as expected given their low-gravity nature. A similar trend has previously been reported by \citet{Martin_2017} down to L0 and \citet{Manjavacas_2020} down to L4, and here we show that the trend maintains also for later L-types for most of the lines. The exception is the K\,I 1516.7\,nm and Na\,I 1139\,nm lines (the two rightmost panels of Fig.~\ref{fig:pEW_spt}), where we see that the distinction between the objects in our sample and the field dwarfs maintains only down to $\sim$L4. Similar behavior for the Na\,I line can be observed in Fig.~21 of \citet{Allers_Liu_2013}. Regarding the K\,I 1516.7\,nm line, field sources with SpT earlier than L5 present larger pEW, whereas young sources present very low pEWs ($<$1.5 $\AA$).  

For the K\,I lines located in the J-band, we also observe that the youngest objects in our sample ($<$30\,Myr; green stars) have on average lower pEW than the objects belonging to the next age group (30$-$100\, Myr; grey pentagons), which is in turn not well distinguished from even older objects in the X-shooter sample. A similar observation has been previously made for the K\,I lines, but also for other lines in the optical portion of the spectra by \citet{Manjavacas_2020}. They found that the pEWs do not significantly change after the age of PLE, meaning that low-mass stars and more massive BDs have
nearly approached their final radii after their first $\sim$100\,Myr, which is also in agreement with the evolutionary models \citep{Baraffe_2015}.
However, although the trend of lower pEWs for young with respect to the field dwarfs seems to maintain throughout the SpT L, we also observe a general drop in pEW values for all objects with SpT in the range L6$-$L8, also seen in plots of \citet{Allers_Liu_2013}. At the transition from L to T-type field dwarfs, we see either a plateau in pEW values or an increase, followed by a decrease towards the later T-types. 

Although we have only a few young T-dwarfs in our sample, it seems that they maintain the trend of lower pEWs at least in two lines (K\,I at 1169.2\,nm and 1177.8\,nm). In other cases, the distinction from the field objects is not that clear. Considering the field age class, the K I doublets in the J-band seem to decrease drastically for later T-types (SpT $\gtrsim$ T4/T5). This behavior was also observed in previous works (e.g, \citealt{Martin_2017}). Clearly, it is an imperative to populate this SpT region with younger T dwarfs to be able to draw further conclusions. However, the considerable decrease of pEW values in field late-T dwarfs might suggest that differences in surface gravity through the K I doublets are no longer distinguishable for late T-types.

It is important to note that the K\,I 1243.7\,nm line is partially blended with a FeH feature that may affect the measurement of the pEWs. Given that the field sources we have used present a lower spectral resolution, their pEWs may artificially have larger values than those measured from the X-shooter spectra. This is the reason why previous studies do not include this line in the analysis \citep{Allers_Liu_2013,Martin_2017}.

Another way to observe the age trends is to plot the pEWs of the alkali lines as a function of the age group, as shown in Fig.~\ref{fig:pEW_age}. For this exercise, we remove all the objects in the T-type range. As expected from Fig.~\ref{fig:pEW_spt}, the pEW on average increases as a function of age (surface gravity), which can be appreciated in the KDEs on the right-hand side of each panel, with the peaks of the distribution shifting upwards with age.

\section{Suggested low-gravity spectral standards}
\label{sect:standards}
There is plenty of spectra of old field dwarfs, with comparable temperatures and SpTs, which can be combined to create spectral templates. On the other hand, spectra of young late-type objects are rare and have a heterogeneous nature in the sense that they encompass a large spread of NIR colors for a fixed optically-based SpT \citep{Allers_Liu_2013,Cruz_2018}. As a result, a well-covered sequence of young LT templates is still lacking. The sequence used for template fitting in this work is that of \citet{Luhman_2017}, covering SpTs L0, L2, L4 and L7, at low spectral resolution. It is important to work towards constructing a more complete sequence that will allow the classification of the new ultracool spectra obtained by JWST and other upcoming facilities. 

With this in mind, in Table \ref{tab:standards_proposed}, we present preliminary spectral standards for spectral classification of low-gravity LT dwarfs that should be used as a starting point towards achieving a more accurate young template sequence. In the selection process, we did not take into account objects in USCO, since these can be affected by extinction, or in PRA, since this association is the oldest in our sample. When more than one spectra seemed suitable as a standard, preference was given to those located in the youngest regions considered in this work. We also prioritized objects with similar SpTs estimated by direct comparison with spectral templates and using spectral type indices, within the respective error range. For some subtypes, there are more than one spectral standards that agree with the selection criteria and thus are proposed. In the end, the standard sequence suggested covers SpTs from L0 down to $\sim$T5 but the subtypes L3, L6, L9, T0, T1, and T4 are lacking since none of the objects in our sample were classified as such.

We note that object \#21 was classified visually as L5, as it appears to be later than L4 and earlier than L7, but closer in shape to the former one. Object \#20 is clearly later than L7, but at the same time lacks T-type characteristics, therefore we proposed the L8 classification. This classification can be re-iterated once a larger number of spectra in this spectral range are known. The object \#53 (T3) was selected as a weak binary candidate in Sect. \ref{subsec:binaries}, but we nevertheless decide to include it here because, even if it is a spectral binary, the components seem to have a very similar spectral type which will not translate into a peculiar unresolved spectrum. The same argument goes to object \#47 which, as already mentioned in Sect. \ref{subsec:binaries}, it is a tight binary where the components have nearly equal luminosity \citep{Best_2017}.

Regarding ages, for L-types, all of the sources selected as possible spectral standards are younger than $\sim$50 Myr. The proposed T3 standard, object \#53, also has comparable age (located in ARG, 40$-$50 Myr). For the remaining two T-type dwarfs, older objects had to be selected: object \#56 located in CARN (200$\pm$50 Myr), and object \#50 in ABDMG (149$^{+51}_{-19}$ Myr). Ideally, one would want a set of templates with ages below 10 Myr to classify the PMO samples that JWST will detect but, at the moment, such data is lacking. What we present here should be seen as a step towards classification of the young late-L to T dwarfs in SFRs that should be revised once new data is acquired.

\begin{table}
\caption{Proposed low-gravity LT spectral standards.}
\centering
\resizebox{0.4\textwidth}{!}{
\begin{tabular}{rllll}
\hline \hline
SpT & ID & Name & Region \\ \hline
L0 & \#33 & J0037-5846 & THA \\
L1 & \#23 & J0045+1634 & ARG  \\
 & \#34 & J0223-5815 & THA \\
 & \#35 & J0323-4631 & THA \\
L2 & \#41 & J0153-6744 & THA  \\
L4 & \#42 & J0342-6817 & THA  \\
L5 & \#21 & J2154-1055 & ARG \\
L7 & \#47$^{a}$ & J1119-1137 & TWA  \\
L8 & \#20 & PSO J318.5-22 & BPMG \\ \hline
T2 & \#56 & J0136+0933 & CARN \\
T3 & \#53$^{b}$ & PSO J168.18−27 & ARG  \\
$\sim$T5 & \#50 & J1110+0116 & ABDMG  \\ \hline
\end{tabular}}
\tablefoot{\footnotesize $^{a}$ Known nearly-equal flux binary \citep{Best_2017}.

$^{b}$ Selected as weak binary candidate in Sect. \ref{subsec:binaries}.}
\label{tab:standards_proposed}
\end{table}

\section{Summary and conclusions}
\label{sec:summary}

We present NIR VLT/X-shooter archival spectra of 56 ultracool dwarfs, probable members of several NYMGs, young clusters and SFRs, with SpTs in the range M8 to T6. The expected ages of the analyzed objects are between 10 and 600\,Myr. We re-determine their SpTs through a comparison with a set of literature young and field templates, and a set of SpT-sensitive indices. With the help of a complementary spectroscopic sample comprising more than 3000 MLT dwarfs of all ages (from nearby SFRs to field), we identify 15 spectral indices that are useful for spectral typing of LT dwarfs and provide their scaling relations with SpT.

As one of the important goals of this paper is to provide means to distinguish young from old LT dwarfs based on their NIR spectra, we identify several spectral indices from the literature suitable for this purpose in the L-type spectral range. In the future, it will be crucial to establish a similar framework for the T-type objects, currently not possible due to the lack of young objects in this spectral range.  
We find that the T-dwarf spectra in our sample consistently show redder colors when compared to their field counterparts, which has been interpreted as a signature of youth. Given their peculiarity when compared to field objects, these spectra also tend to be recognized as potential unresolved spectral binaries. We find the latter interpretation unlikely, given the low prevalence of binaries in high spatial resolution surveys of T-dwarfs, and conclude that the red colors are probably caused by the effects of low surface gravity.

We also inspected the behavior of several alkali lines in the $J-$ and the $H-$bands, observing the correlations of their widths with the SpT and age. The pEWs of the K\,I lines in the $J-$band typically increase from M8 to $\sim$L4, displaying a turnover towards later L-types. The K\,I line in the $H-$band (1516.7\,nm) and the Na\,I $J-$band line (1139\,nm) show a constant decrease in pEWs with increasing SpT. These observations are particularly valid for the field dwarfs, while the pEWs of the objects in our sample which are significantly younger have lower values, and do not show strong trends among them. The pEWs of the alkali lines show clear trends with age, as a consequence of changes in the surface gravity. 

Finally, we identify 12 objects as preliminary spectral standards for young L and T dwarfs, which should be useful for spectral typing of the objects identified with the new facilities, such as the JWST. More high-quality spectra of young ultracool dwarfs, in particular those with SpTs in the late L- and the T-type regime are needed to establish a complete set of templates for the characterization of young brown dwarfs and PMOs. 

\begin{acknowledgements}
We are grateful to Carlo Manara for providing the code to calculate the pseudo-equivalent widths of the alkali lines. L.P., K.M., and V.A. acknowledge funding by the
Science and Technology Foundation of Portugal (FCT), grants PTDC/FISAST/7002/2020, UIDB/00099/2020, 
PTDC/FIS-AST/28731/2017, IF/00194/2015, SFRH/BD/143433/2019, 2022.03809.CEECIND, UIDB/04434/2020 and UIDP/04434/2020.
L.P. acknowledges financial support from the CSIC project JAEICU-21-ICE-09, from Centro Superior de Investigaciones Cient\'{i}ficas (CSIC) under the PIE project 20215AT016, and the program Unidad de Excelencia Mar\'{i}a de Maeztu CEX2020-001058-M.
This work has made use of data from the European Space Agency (ESA) mission {\it Gaia} (\url{https://www.cosmos.esa.int/gaia}), processed by the {\it Gaia} Data Processing and Analysis Consortium (DPAC, \url{https://www.cosmos.esa.int/web/gaia/dpac/consortium}). Funding for the DPAC has been provided by national institutions, in particular the institutions participating in the {\it Gaia} Multilateral Agreement. 
This research has benefitted from the SpeX Prism Spectral Libraries, maintained by Adam Burgasser at \url{http://www.browndwarfs.org/spexprism}. This research has benefitted from the Montreal Brown Dwarf and Exoplanet Spectral Library, maintained by Jonathan Gagn\'{e}.
\end{acknowledgements}

\bibliographystyle{aa}
\bibliography{xshoo_spec}

\begin{appendix} 

\section{Additional tables}

In Table \ref{tab:xshoo_observations} we summarize basic observational parameters of our sample presented in Sect. \ref{sec:data}.

Table \ref{tab:spec_class} shows the spectral classification results discussed in Sect. \ref{sec:spt_ext}.

\renewcommand{\arraystretch}{1.2}

\begin{table*}
\caption{The objects in our dataset, along with some basic observational parameters.}
\centering
\resizebox{0.80\textwidth}{!}{
\begin{tabular}{llllllllll}
\hline \hline
\multirow{2}{*}{ID} & \multirow{2}{*}{Name} & \multirow{2}{*}{RA} & \multirow{2}{*}{DEC} & \multirow{2}{*}{Date observation} & \multirow{2}{*}{ESO PROG ID} & \multirow{2}{*}{EXPTIME} & \multirow{2}{*}{NEXP} & \multirow{2}{*}{Slit ($''$)} & \multirow{2}{*}{R ($\lambda/\Delta\lambda$)$^{a}$} \\
 &  & \multicolumn{1}{c}{} &  &  &  &  &  &  \\ \hline
2 & UScoJ160606-233513 & 16:06:06.29 & -23:35:13.3 & 2014-07-04 & 093.C-0769(A) & 197 & 16 & 1.2 & 4300 \\
3 & UScoJ160723-221102 & 16:07:23.82 & -22:11:02.0 & 2014-04-22 & 093.C-0769(A) & 197 & 8 & 1.2 & 4300 \\
 &  &  &  & 2016-07-25 & 097.C-0592(A) & 290 & 8 & 0.9 & 5600 \\
4 & UScoJ160737-224247 & 16:07:37.99 & -22:42:47.0 & 2014-07-04 & 093.C-0769(A) & 197 & 14 & 1.2 & 4300 \\
 &  &  &  & 2014-08-02 & 093.C-0769(A) & 197 & 32 & 1.2 & 4300 \\ 
5 & UScoJ160818-223225 & 16:08:18.43 & -22:32:25.0 & 2014-07-02 & 093.C-0769(A) & 197 & 28 & 1.2 & 4300 \\
 &  &  &  & 2014-07-03 & 093.C-0769(A) & 197 & 14 & 1.2 & 4300 \\
 &  &  &  & 2014-07-04 & 093.C-0769(A) & 197 & 14 & 1.2 & 4300 \\ 
6 & UScoJ160828-231510 & 16:08:28.47 & -23:15:10.4 & 2014-06-15 & 093.C-0769(A) & 197 & 14 & 1.2 & 4300 \\
 &  &  &  & 2014-06-20 & 093.C-0769(A) & 197 & 14 & 1.2 & 4300 \\ 
7 & UScoJ161302-212428 & 16:13:02.32 & -21:24:28.5 & 2015-04-13 & 095.C-0812(A) & 300 & 4 & 1.2 & 4300 \\ 
8 & UScoJ160918-222923 & 16:09:18.69 & -22:29:23.9 & 2015-04-13 & 095.C-0812(A) & 300 & 16 & 1.2 & 4300 \\
9 & VISTAJ1559−2214 & 15:59:36.38 & -22:14:15.9 & 2015-04-13 & 095.C-0812(A) & 300 & 4 & 1.2 & 4300 \\ 
10 & VISTAJ1614−2331 & 16:14:22.56 & -23:31:17.8 & 2015-04-14 & 095.C-0812(A) & 300 & 6 & 1.2 & 4300 \\ 
11 & VISTAJ1611−2215 & 16:11:44.37 & -22:15:44.6 & 2015-04-12 & 095.C-0812(A) & 300 & 8 & 1.2 & 4300 \\ 
12 & VISTAJ1607−2146 & 16:07:31.61 & -21:46:54.4 & 2015-04-14 & 095.C-0812(A) & 300 & 8 & 1.2 & 4300 \\ 
13 & VISTAJ1604−2241 & 16:04:13.04 & -22:41:03.4 & 2015-04-13 & 095.C-0812(A) & 300 & 8 & 1.2 & 4300 \\ 
14 & VISTAJ1614−2211 & 16:14:07.56 & -22:11:52.2 & 2015-04-12 & 095.C-0812(A) & 300 & 10 & 1.2 & 4300 \\ 
15 & VISTAJ1605−2403 & 16:05:39.09 & -24:03:32.8 & 2015-04-12 & 095.C-0812(A) & 300 & 10 & 1.2 & 4300 \\ 
16 & VISTAJ1604−2134 & 16:04:20.04 & -21:34:53.0 & 2015-04-13 & 095.C-0812(A) & 300 & 10 & 1.2 & 4300 \\ 
17 & VISTAJ1601−2212 & 16:01:36.92 & -22:12:02.7 & 2015-04-13 & 095.C-0812(A) & 300 & 10 & 1.2 & 4300 \\ 
18 & VISTAJ1615−2229 & 16:15:12.70 & -22:29:49.2 & 2015-04-14 & 095.C-0812(A) & 300 & 10 & 1.2 & 4300 \\ 
19 & J0032-4405 & 00:32:55.80 & -44:05:05.8 & 2016-07-16 & 097.C-0592(A) & 300 & 6 & 0.9 & 5600 \\
20 & PSO J318.5-22 & 21:14:08.02 & -22:51:35.8 & 2018-06-20 & 0101.C-0290(A) & 300 & 12 & 0.6 & 8100 \\
 &  &  &  & 2018-06-27 & 0101.C-0290(A) & 300 & 12 & 0.6 & 8100 \\ 
21 & J2154-1055 & 21:54:34.50 & -10:55:30.8 & 2013-08-22 & 091.C-0462(B) & 290 & 10 & 0.9 & 5600 \\ 
23 & J0045+1634 & 00:45:21.42 & 16:34:44.7 & 2018-10-09 & 0102.C-0121(A) & 234 & 12 & 0.6 & 8100 \\ 
24 & J0518-2756 & 05:18:46.20 & -27:56:45.8 & 2018-11-27 & 0102.C-0121(A) & 234 & 12 & 0.6 & 8100 \\ 
25 & 2MASS J03552337+1133437 & 03:55:23.60 & 11:33:34.3 & 2019-09-10 & 0103.C-0231(A) & 175 & 12 & 0.6 & 8100 \\ 
26 & 2MASS J14252798-3650229 & 14:25:27.98 & -36:50:23.0 & 2018-06-19 & 0101.C-0290(A) & 234 & 12 & 0.6 & 8100 \\ 
27 & J0339-3525 (LP 944-20) & 03:39:35.25 & -35:25:43.6 & 2013-10-18 & 092.D-0128(A) & 225 & 20 & 0.4 & 11600 \\ 
28 & 2MASS J03264225-2102057 & 03:26:42.25 & -21:02:05.8 & 2018-10-21 & 0102.C-0121(A) & 234 & 12 & 0.6 & 8100 \\ 
29 & 2MASS J22064498-4217208 & 22:06:44.98 & -42:17:21.1 & 2013-08-22 & 091.C-0462(B) & 290 & 6 & 0.9 & 5600 \\ 
31 & J2322-6151 & 23:22:53.00 & -61:51:27.5 & 2018-06-26 & 0101.C-0290(A) & 234 & 12 & 0.6 & 8100 \\ 
32 & J0006-6436 & 00:06:57.93 & -64:36:54.2 & 2016-07-22 & 097.C-0592(A) & 300 & 4 & 0.9 & 5600 \\ 
33 & J0037-5846 & 00:37:43.06 & -58:46:22.8 & 2016-08-11 & 097.C-0592(A) & 290 & 8 & 0.9 & 5600 \\
 &  &  &  & 2013-08-22 & 091.C-0462(B) & 290 & 10 & 0.9 & 5600 \\ 
34 & J0223-5815 & 02:23:54.65 & -58:15:06.6 & 2016-08-23 & 097.C-0592(A) & 300 & 6 & 0.9 & 5600 \\
 &  &  &  & 2016-08-26 & 097.C-0592(A) & 300 & 6 & 0.9 & 5600 \\ 
35 & J0323-4631 & 03:23:10.01 & -46:31:23.6 & 2016-08-10 & 097.C-0592(A) & 290 & 8 & 0.9 & 5600 \\
 &  &  &  & 2013-08-22 & 091.C-0462(B) & 290 & 6 & 0.9 & 5600 \\ 
36 & J0141-4633 & 01:41:58.23 & -46:33:57.3 & 2016-08-20 & 097.C-0592(A) & 300 & 6 & 0.9 & 5600 \\ 
37 & J0357-4417 & 03:57:26.96 & -44:17:30.5 & 2013-08-22 & 091.C-0462(B) & 290 & 4 & 0.9 & 5600 \\ 
38 & J0117-3403 & 01:17:47.49 & -34:03:25.8 & 2016-08-18 & 097.C-0592(A) & 290 & 8 & 0.9 & 5600 \\ 
39 & J0241-0326 & 02:41:11.51 & -03:26:58.8 & 2013-08-22 & 091.C-0462(B) & 290 & 8 & 0.9 & 5600 \\ 
40 & J0004-6410 & 00:04:02.88 & -64:10:35.8 & 2013-08-22 & 091.C-0462(B) & 290 & 10 & 0.9 & 5600 \\ 
41 & 2MASS J01531463-6744181 & 01:53:14.63 & -67:44:18.1 & 2018-10-17 & 0102.C-0121(A) & 234 & 12 & 0.6 & 8100 \\
 &  &  &  & 2018-10-18 & 0102.C-0121(A) & 234 & 12 & 0.6 & 8100 \\ 
42 & 2MASS J03421621-6817321 & 03:42:16.21 & -68:17:32.1 & 2018-10-08 & 0102.C-0121(A) & 234 & 24 & 0.6 & 8100 \\

46 & J1207-3900 & 12:07:48.35 & -39:00:04.5 & 2015-05-31 & 095.C-0147(A) & 900 & 4 & 1.2 & 4300 \\
 &  &  &  & 2015-06-01 & 095.C-0147(A) & 900 & 4 & 1.2 & 4300 \\
 &  &  &  & 2018-04-28 & 0101.C-0290(A) & 234 & 12 & 0.6 & 8100 \\ 
47 & 2MASS J11193254-1137466 & 11:19:32.54 & -11:37:46.6 & 2019-04-15 & 0103.C-0231(A) & 300 & 12 & 0.6 & 8100 \\ 
48 & 2MASS J114724.10-204021.3 & 11:47:24.10 & -20:40:21.3 & 2018-06-16 & 0101.C-0290(A) & 300 & 12 & 0.6 & 8100 \\
49 & WISE J052857.69+090104.2 & 05:28:57.68 & 09:01:04.4 & 2018-10-22 & 0102.C-0121(A) & 234 & 12 & 0.6 & 8100 \\
 &  &  &  & 2018-10-26 & 0102.C-0121(A) & 234 & 24 & 0.6 & 8100 \\
50 & SDSSp J111010.01+011613.1 & 11:10:10.01 & 01:16:13.1 & 2018-04-29 & 0101.C-0290(A) & 234 & 12 & 0.6 & 8100 \\
 &  &  &  & 2018-05-27 & 0101.C-0290(A) & 234 & 12 & 0.6 & 8100 \\
 &  &  &  & 2018-05-28 & 0101.C-0290(A) & 234 & 12 & 0.6 & 8100\\ 
51 & SDSS J152103.24+013142.7 & 15:21:03.27 & 01:31:42.7 & 2011-06-09 & 087.C-0639(A) & 290 & 4 & 0.9 & 5600 \\
52 & ULAS J004757.41+154641.4 & 00:47:57.43 & 15:46:41.2 & 2011-09-21 & 087.C-0639(B) & 490 & 4 & 0.9 & 5600 \\ 
53 & PSO J168.1800−27.2264 & 11:12:43.25 & -27:13:36.1 & 2014-03-12 & 092.C-0229(B) & 600 & 6 & 1.2 & 4300 \\ 
55 & ULAS J131610.13+031205.5 & 13:16:10.13 & 03:12:05.6 & 2011-06-06 & 087.C-0639(A) & 390 & 4 & 0.9 & 5600 \\
56 & SIMP J013656.5+093347.3 & 01:36:56.56 & 09:33:47.3 & 2018-11-16 & 0102.C-0121(A) & 234 & 12 & 0.6 & 8100 \\ 
57 & 2MASS J08370450+2016033 & 08:37:04.49 & 20:16:03.2 & 2017-01-01 & 098.C-0277(A) & 300 & 8 & 1.2 & 4300 \\ 
58 & UGCS J084510.65+214817.0 & 08:45:10.66 & 21:48:17.1 & 2017-01-02 & 098.C-0277(A) & 300 & 10 & 1.2 & 4300 \\ 
59 & 2MASS J03463425+2350036 & 03:46:34.25 & 23:50:03.6 & 2016-11-11 & 098.C-0277(A) & 300 & 10 & 1.2 & 4300 \\ 
60 & 2MASS J03541027+2341402 & 03:54:10.27 & 23:41:40.2 & 2016-11-03 & 098.C-0277(A) & 300 & 14 & 1.2 & 4300 \\ 
61 & 2MASS J00464841+0715177 & 00:46:48.42 & 07:15:17.8 & 2019-07-15 & 0103.C-0231(A) & 175 & 12 & 0.6 & 8100 \\
62 & 2MASS J05120636-2949540 & 05:12:06.37 & -29:49:54.0 & 2018-10-31 & 0102.C-0121(A) & 234 & 12 & 0.6 & 8100 \\
63 & Gu Psc b & 01:12:36.54 & 17:04:29.9 & 2018-11-14 & 0102.C-0121(A) & 234 & 12 & 0.6 & 8100 \\
 &  &  &  & 2018-11-25 & 0102.C-0121(A) & 234 & 12 & 0.6 & 8100 \\
 &  &  &  & 2018-11-26 & 0102.C-0121(A) & 234 & 12 & 0.6 & 8100  \\ \hline
\end{tabular}}
\tablefoot{\tiny{$^{a}$ Spectral resolutions as presented in the X-shooter's Characteristics Webpage (\url{https://www.eso.org/sci/facilities/paranal/instruments/xshooter/inst.html}) depending on wavelength and slit width.}}
\label{tab:xshoo_observations}
\end{table*}

\begin{table*}
\centering
    \caption{Spectral classification results.}
    \resizebox{0.99\textwidth}{!}{
\begin{tabular}{lllllllll}
\hline \hline
 ID & Name & Region & SpT Literature & References$^{a}$ & SpT Templates$^{f}$ & A$_{V}$ & Templates Set & SpT Indices \\
\hline
2 & UScoJ160606-233513 & USCO & M8.5$\pm$0.5 & \citet{Luhman_2018}$^{e}$ & M8.5$\pm$0.5 & 1.4 & Young & M9.5$\pm$0.5  \\
3 & UScoJ160723-221102 & USCO & M8.5$\pm$0.5 & \citet{Luhman_2018} & M7.0$\pm$0.5 & 1.4 & Young & M8.5$\pm$0.5  \\
4 & UScoJ160737-224247 & USCO & M9.0$\pm$0.5 & \citet{Luhman_2018} & M9.0$\pm$0.5 & 1.6 & Young & L0.5$\pm$0.5  \\
5 & UScoJ160818-223225 & USCO & M9.25$\pm$0.5 & \citet{Luhman_2018} & M9.5$\pm$0.5 & 0.0 & Young & L1.0$\pm$0.5 \\
6 & UScoJ160828-231510 & USCO & M9.0$\pm$0.5 & \citet{Luhman_2018} & M9.0$\pm$0.5 & 1.0 & Young & M9.0$\pm$0.5  \\
7 & UScoJ161302-212428 & USCO & M9.5$\pm$0.5 & \citet{Luhman_2018} & M9.0$\pm$0.5 & 1.8 & Young & L0.0$\pm$1.0 \\
8 & UScoJ160918-222923 & USCO & L1.0$\pm$0.5 & \citet{Luhman_2018} & L2.0$\pm$2.0 & 0.0 & Young & L2.0$\pm$0.5  \\
9 & VISTAJ1559−2214 & USCO & M9.5$\pm$0.5 & \citet{Luhman_2018} & M9.5$\pm$0.5 & 0.0 & Young & M9.5$\pm$1.0  \\
10 & VISTAJ1614−2331 & USCO & M9.5$\pm$0.5 & \citet{Luhman_2018} & M9.5$\pm$0.5 & 0.4 & Young & L1.5$\pm$1.0 \\
11 & VISTAJ1611−2215 & USCO & L3.0$\pm$0.5 & \citet{Luhman_2018} & L4.0$^{+3}_{-2}$ & 0.6 & Young & M9.0$\pm$1.0 \\ 
12 & VISTAJ1607−2146 & USCO & L1.0$\pm$0.5 & \citet{Luhman_2018} & L0.0$\pm$2.0 & 1.6 & Young & L1.5$\pm$1.0  \\
13 & VISTAJ1604−2241 & USCO & L1.0$\pm$0.5 & \citet{Luhman_2018} & L2.0$\pm$2.0 & 1.0 & Young & M9.5$\pm$1.0 \\
14 & VISTAJ1614−2211 & USCO & L1.0$\pm$0.5 & \citet{Luhman_2018} & L0.0$\pm$2.0 & 1.4 & Young & M9.5$\pm$0.5  \\
15 & VISTAJ1605−2403 & USCO & L1.0$\pm$0.5 & \citet{Luhman_2018} & L2.0$\pm$2.0 & 0.4 & Young & L1.5$\pm$0.5  \\
16 & VISTAJ1604−2134 & USCO & L6.0$\pm$1.0 & \citet{Lodieu_2018} & L2.0$\pm$2.0 & 0.4 & Young & L1.0$\pm$0.5  \\
17 & VISTAJ1601−2212 & USCO & L4.0$\pm$0.5 & \citet{Luhman_2018} & L4.0$^{+3}_{-2}$ & 1.6 & Young & L2.0$\pm$1.0   \\
18 & VISTAJ1615−2229 & USCO & L4.0$\pm$0.5 & \citet{Luhman_2018} & L4.0$^{+3}_{-2}$ & 1.2 & Young & L3.0$\pm$1.0  \\
19 & J0032-4405 & ABDMG & L0.0$\pm$1.0 & \citet{Ujjwal_2020}; \citet{Allers_Liu_2013} & L0.0$\pm$2.0 &   & Young & L1.0$\pm$0.5  \\
20 & PSO J318.5-22 & BPMG & L7.0$\pm$1.0 & \citet{Liu_2013_obj20} & L8.0$\pm$1.0 &   & Young & L4.0$\pm$3.0   \\
21 & J2154-1055 & ARG & L5.0 $\beta$/$\gamma$ & \citet{Gagne_2014_obj21,Gagne_2015_VII} & L5.0$\pm$1.0 &   & Young & L4.5$\pm$3.0  \\
23 & J0045+1634 & ARG & L2.0$\pm$1.0 & \citet{Gagne_2014,Allers_Liu_2013} & L1.0$\pm$1.0 &   & Young & L1.5$\pm$1.0   \\
24 & J0518-2756 & COL & L1.0$\pm$1.0 & \citet{Gagne_2014,Allers_Liu_2013} & L1.0$\pm$1.0 &   & Young & L1.0$\pm$0.5   \\
25 & 2MASS J03552337+1133437 & ABDMG & L3.0$\pm$1.0 & \citet{Liu_2013,Allers_Liu_2013} & L7.0$\pm$3.0 &   & Young & L2.5$\pm$1.0   \\
26 & 2MASS J14252798-3650229 & ABDMG & L4.0 $\gamma$ & \citet{Gagne_2015_VII} & L2.0$\pm$2.0 &   & Young & L4.0$\pm$0.5   \\
27 & J0339-3525 & CAS & L0.0$\pm$1.0 & \citet{Ribas_2003}, \citet{Gagne_2014}; \citet{Allers_Liu_2013} & M9.0$\pm$1.0 &   & Field & M9.5$\pm$0.5   \\
28 & 2MASS J03264225-2102057 & ABDMG & L5.0$\beta \gamma$ & \citet{Faherty_2016} & L4.0$^{+3}_{-2}$ &   & Young & L5.0$\pm$3.5  \\
29 & 2MASS J22064498-4217208 & ABDMG & L4.0 $\gamma$ & \citet{Faherty_2016} & L2.0$\pm$2.0 &   & Young & L3.0$\pm$0.5  \\
31 & J2322-6151 & THA & L3.0 $\gamma$ & \citet{Gagne_2014,Faherty_2016} & L1.0$\pm$1.0 &   & Young & L2.5$\pm$0.5   \\
32 & J0006-6436 & THA & M8.0 $\gamma$ & \citet{Gagne_2014,Gagne_2015_VII} & M9.0$\pm$0.5 &   & Young & L0.0$\pm$0.5  \\
33 & J0037-5846 & THA & L0.0 $\gamma ^{b}$ & \citet{Gagne_2014,Cruz_2009} & L0.0$\pm$2.0 &   & Young & L1.0$\pm$1.0  \\
34 & J0223-5815 & THA & L0.0 $\gamma ^{b}$ & \citet{Gagne_2014,Cruz_2009} & L1.0$\pm$1.0 &   & Young & L0.5$\pm$0.5  \\
35 & J0323-4631 & THA & L0.0 $\gamma$ & \citet{Gagne_2014,Faherty_2016} & L1.0$\pm$1.0 &   & Young & L0.5$\pm$0.5  \\
36 & J0141-4633 & THA & L0.0$\pm$1.0 & \citet{Gagne_2014,Allers_Liu_2013} & L0.0$\pm$2.0 &   & Young & L1.5$\pm$1.0 \\
37 & J0357-4417 & THA & L2.0$\pm$0.5 pec$^{c}$ & \citet{Gagne_2014,Marocco_2013} & M9.5$\pm$0.5 &   & Young & L0.5$\pm$0.5  \\
38 & J0117-3403 & THA & L1.0$\pm$1.0 & \citet{Gagne_2014,Allers_Liu_2013} & L0.0$\pm$2.0 &   & Young & L2.5$\pm$0.5  \\
39 & J0241-0326 & THA & L1.0$\pm$1.0 & \citet{Gagne_2014,Allers_Liu_2013} & L0.0$\pm$2.0 &   & Young & L1.0$\pm$0.5  \\
40 & J0004-6410 & THA & L1.0 $\gamma$ & \citet{Kirkpatrick_2010,Faherty_2016} & L2.0$\pm$2.0 &   & Young & L2.0$\pm$1.0  \\
41 & 2MASS J01531463-6744181 & THA & L3.0 $\beta$ & \citet{Faherty_2016} & L2.0$\pm$2.0 &   & Young & L3.5$\pm$0.5  \\
42 & 2MASS J03421621-6817321 & THA & L4.0$\gamma ^{b}$ & \citet{Gagne_2014,Gagne_2015_VII} & L4.0$^{+3}_{-2}$ &   & Young & L3.0$\pm$1.0 \\
46 & J1207-3900 & TWA & L1.0$\pm$1.0 & \citet{Gagne_2014_obj46} & M9.5$\pm$0.5 &   & Young & L1.0$\pm$0.5  \\
47 & 2MASS J11193254-1137466 & TWA & L7.0 & \citet{Kellog_2016} & L7.0$\pm$3.0 &   & Young & L4.5$\pm$3.0   \\
48 & 2MASS J114724.10-204021.3 & TWA & L7.0$\pm$1.0 & \citet{Schneider_2016} & L7.0$\pm$3.0 &   & Young & L6.0$\pm$3.5 \\
49 & WISE J052857.69+090104.2 & 32OR & L1.0 & \citet{Burgasser_2016} & L0.0$\pm$2.0 &   & Young & L1.0$\pm$0.5  \\
50 & SDSSp J111010.01+011613.1 & ABDMG & T5.5 & \citet{Gagne_2015_obj50,Burgasser_2006} & T5.0$\pm$1.0 &   & Field & T5.5$\pm$0.5  \\
51 & SDSS J152103.24+013142.7 & ARG & T2.0$\pm$0.5 & \citet{Zhang_2021,Knapp_2004} & T3.0$\pm$1.0 &   & Field & T2.5$\pm$0.5   \\

53 & PSO J168.1800−27.2264 & ARG & T2.5$\pm$0.5 & \citet{Zhang_2021,Best_2015} & T3.0$\pm$1.0 &   & Field & T3.0$\pm$0.5   \\

56 & SIMP J013656.5+093347.3 & CARN & T2.5$\pm$0.5 & \citet{Gagne_2017_obj56,Artigau_2006} & T2.0$\pm$1.0 &   & Field & T2.0$\pm$0.5  \\
57 & 2MASS J08370450+2016033 & PRA & L0.0 & \citet{Manjavacas_2020} & L1.0$\pm$1.0 &   & Field & M9.5$\pm$0.5   \\
58 & UGCS J084510.65+214817.0 & PRA & L1.5 & \citet{Manjavacas_2020} & L1.0$\pm$1.0 &   & Field & L0.0$\pm$0.5   \\
59 & 2MASS J03463425+2350036 & PLE & L1.0 & \citet{Manjavacas_2020} & L1.0$\pm$1.0 &   & Field & L1.0$\pm$0.5   \\
60 & 2MASS J03541027+2341402 & PLE & L3.0 & \citet{Manjavacas_2020} & L2.0$\pm$1.0 &   & Field & L2.5$\pm$0.5  \\
61 & 2MASS J00464841+0715177 & BPMG & L0.0 $\delta$ & \citet{Gagne_2015_VII} & M9.5$\pm$0.5 &   & Young & L1.0$\pm$0.5   \\
62 & 2MASS J05120636-2949540 & BPMG$^{d}$ & L5.0 $\beta$ & \citet{Gagne_2015_VII} & L4.0$^{+3}_{-2}$ &   & Young & L5.0$\pm$3.5  \\ 
63 & Gu Psc b & ABDMG & T3.5$\pm$1.0 & \citet{Naud_2014} & T3.0$\pm$1.0 &   & Field & T3.0$\pm$0.5   \\
\hline
\end{tabular}}
\tablefoot{\footnotesize $^{a}$ The first reference is the membership (region) reference whilst the second is the SpT reference. If there is only one reference it means that both the region and SpT information were retrieved from that work.

$^{b}$ Literature SpT based on optical data.

$^{c}$ Known unresolved binary, the SpT given is the unresolved classification.

$^{d}$ Uncertain membership. Object discussed in Appendix \ref{app:specific_objs}.

$^{e}$ Combined spectral type, derived from a mixture of their own classification (optical or NIR) and literature.

$^{f}$ Final SpT classification adopted throughout the analysis (see Sect. \ref{subsec:spt_temp}).
}
\label{tab:spec_class}
\end{table*}

\section{X-shooter spectra}\label{app:other_spec}

\begin{figure}[h]
\centering
    \includegraphics[width=0.49\textwidth]{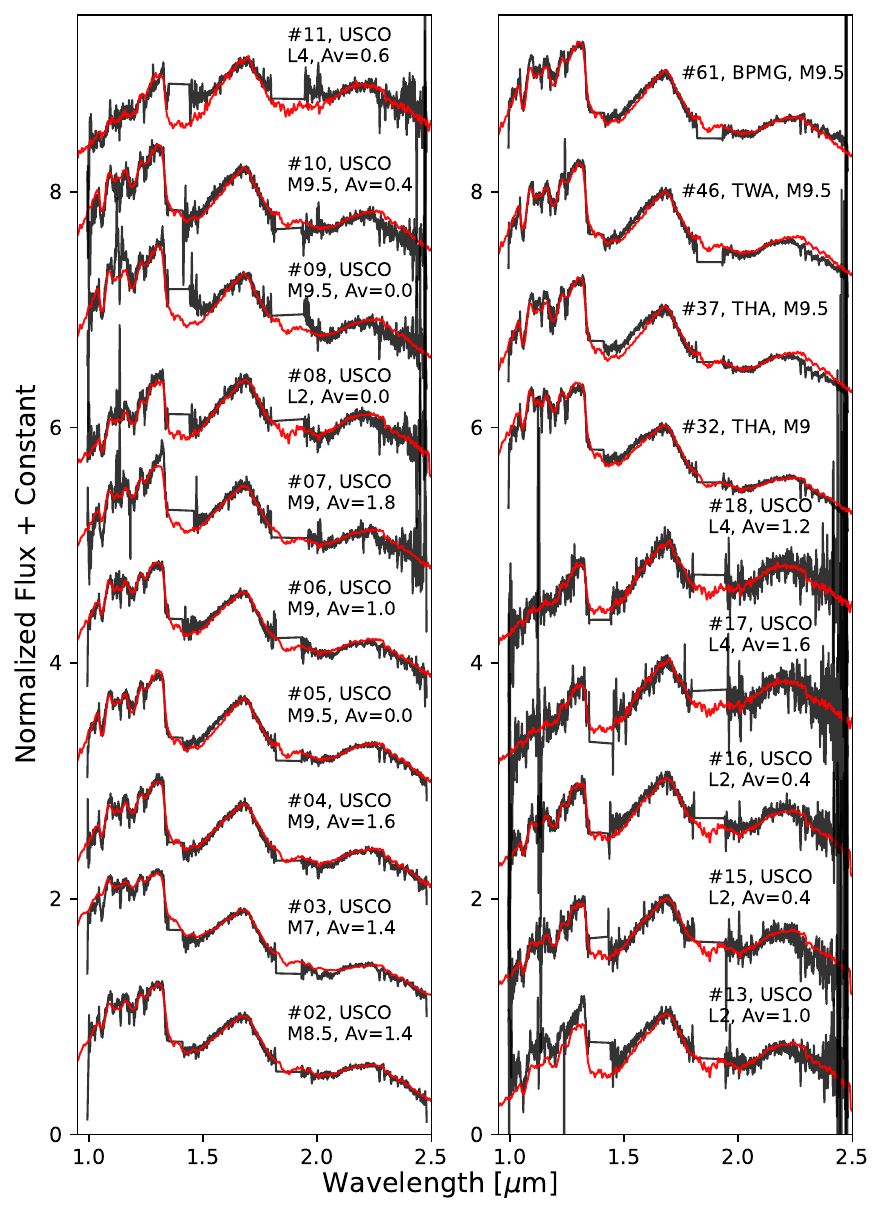}
\caption{Remaining spectra of our sample along their best-fitted template in red (see Sect. \ref{subsec:spt_temp}). For all of the objects presented here, the best-fit was accomplished with a young template. The ID, young association, final SpT and extinction derived (for objects in USCO) are shown above the correspondent spectrum (in black). All spectra are normalized at 1.67 $\mu$m. An offset between the spectra was added for clarity.}
\label{fig_app:other_spec}       
\end{figure}

In Fig. \ref{fig_app:other_spec} we show the remaining X-shooter spectra not present in the paper, along with their best-fitted spectral template (see Sect. \ref{subsec:spt_temp}). All the retrieved best-fits were accomplished with young templates (in red).

\section{Discussion on specific objects}\label{app:specific_objs}

In this section, we present a short discussion on the membership of some specific objects.
To do that, we use the Bayesian Analysis for Nearby Young Associations (BANYAN) $\Sigma$, a tool that determines the membership probability of an object belonging to a nearby young association \citep{Gagne_2018_BANYAN}. It requests as mandatory inputs the coordinates (RA and DEC, in decimal degrees), the proper motion (in mas/yr), and respective errors, both in right ascension (RA) and declination (DEC). Optionally, the radial velocity (in km/s) and parallax (in mas) of the objects, along with the associated errors, can also be given. If such information is lacking, the membership probability will still be computed and the algorithm will estimate the most likely distance to the object and the radial velocity that it should have if it was a member of a specific nearby young association.

\subsection{Object \#19 (J0032-4405)}

As stated by \citet{Dupuy_2018}, object \#19 has two parallax values in the literature: the one from \citet{Faherty_2012} which was considered in \citet{Gagne_2014} and results in membership to BPMG, and the one from \citet{Faherty_2016} which leads to ABDMG.

In the color-absolute magnitude diagram (see Fig.~\ref{fig:cmd}), object \#19 is in-between both the young and the field sequences although it appears to be more in agreement with the field sequence. When comparing with spectral templates, the best fit was initially achieved with an L2 field template (see Sect.~\ref{subsec:spt_temp}). However, after visual inspection (see Fig.~\ref{fig:field_vs_young}), we maintained the classification obtained by the best young template (L0$\pm$2) since it appeared to better reproduce the K-band. Nonetheless, we recognize that a subtle "shoulder" at $\sim$1.57$\mu$m is noticeable, which is typical of intermediate-gravity objects \citep{Allers_Liu_2013} and thus suggests a slightly older age than some other objects of the sample. Moreover, in the TLI-g diagram in Fig.~\ref{fig:gravsens_indices} object \#19 is clearly in agreement with the mid-gravity age class of AA22 if we consider its SpT to be L0 (considering a young template) but even more so if we consider its SpT to be L2 (first estimate obtained with the best-fit being a field template in Sect.~\ref{subsec:spt_temp}). Finally, its alkali lines are broader than those of the age class $<$30\,Myr which is where the objects located in BPMG are included. Hence, all points towards a membership to ABDMG.

To re-analyze the nature of object \#19, we implemented the BANYAN $\Sigma$ tool considering the parallax and proper motion values from \textit{Gaia} DR3 ($\pi = 28.9542\pm0.4217 ''$, $\mu_{\alpha} = 127.841\pm0.284$ mas/yr, $\mu_{\delta} = -96.832\pm0.314$ mas/yr) and the radial velocity of \citet{Faherty_2016} ($r_{V} = 12.95\pm1.92$ km/s). The result corroborates a membership to ABDMG ($\sim$150\,Myr) with a probability of 99.4\%, where the remaining 0.6\% suggest the field population. This result is in agreement with previous works such as \citet{Ujjwal_2020}. 

\subsection{Object \#62 (2MASS J05120636-2949540)}

The derived SpT for object \#62 was L4$^{+3}_{-2}$, considering a young template (see Sect.~\ref{subsec:spt_temp}). Throughout this work, we considered that object \#62 was a member of BPMG (age 24$\pm$2\,Myr, \citealt{bell_2015}). However, it shows high pEWs values for the alkali lines compared to the remaining objects within the $<$30\,Myr age class. Its values seem to be more in agreement with those of the objects located in the age class 100$-$300\,Myr. It is important to state that this object's spectrum has an artifact in the range of calculation of the 1243.7\,nm K\,I line,
therefore, it was not included in the panels corresponding to this line in Figs.~\ref{fig:pEW_spt} and \ref{fig:pEW_age}. 

Regarding the gravity-sensitive indices, object \#62 has not a trivial behavior. In some indices, such as HPI, it has a value that is in agreement with the mid-gravity class, towards the young sequence.  On the other hand, its TLI-g and CH4-K index values still agree better with the intermediate-gravity age class but this time towards the overlapping region with the field objects of AA22. This behavior might suggest that object \#62 does not belong to BPMG, which is among the youngest NYMGs in our sample.

We implemented the BANYAN $\Sigma$ tool considering: $\mu_{\alpha} = -5.0\pm5.1$ mas/yr, $\mu_{\delta} = 84.5\pm6.6$ mas/yr, $r_{v} = 19.8\pm1.5$ km/s \citep{Gagne_2015_V}, and $\pi = 49.6\pm2.8 ''$ \citep{Best_2021}. The algorithm presented a membership probability that completely favors the field population. 
We conclude that object \#62 might not be as young as members of BPMG. But since it exhibits signs of youth (e.g, triangular H-band), it does not seem to be as old as objects from the field population. Therefore, its membership is uncertain; it may belong to an yet undefined young association, or be an isolated object that was formed in a protoplanetary disk and ejected early in its evolution.

\section{NIR spectral indices}\label{app:indices}

In Table \ref{tab_app:spec_indices} we summarize the functional forms of all the SpT indices inspected in Sect.~\ref{subsec:spt_indices} along with the respective references and targeted spectral features. Some of these indices demonstrated a gravity-sensitive behavior in a specific SpT range, but they were all defined in the literature to be spectral indices that correlate well with SpT for a specific range initially designated.

There are spectral indices in the literature that were, in fact, defined to be gravity-sensitive, and the information regarding those is presented in Table \ref{tab_app:youth_indices}.

\onecolumn
\begin{longtable}{lccc}
\caption{Spectral type indices inspected.}\\ 

\hline\hline
Index & Reference & Formula & Feature Measured\\
\hline
\endfirsthead

\caption{continued.}\\
\hline\hline
Index & Reference & Formula & Feature Measured \\
\hline
\endhead

\hline
\endfoot
            H2O & (1) & $\frac{\langle F_{\lambda=1.550-1.560}\rangle}{\langle F_{\lambda=1.492-1.502} \rangle} $ & $\sim$1.50-1.57 $\mu$m H$_{2}$O absorption \\
            TLI-K & (2) & $\frac{\langle F_{\lambda=1.97-1.99}\rangle}{\langle F_{\lambda=2.22-2.24} \rangle} $ & $\sim$1.97 $\mu$m Ca lines \\
            TLI-J & (2) & $\frac{\langle F_{\lambda=1.20-1.21}\rangle}{\langle F_{\lambda=1.27-1.29} \rangle} $ & $\sim$1.2 $\mu$m FeH absorption band\\
            sHJ & (3) & $\frac{\langle F_{\lambda=1.265-1.305}\rangle - \langle F_{\lambda=1.6-1.7} \rangle}{0.5(\langle _{\lambda=1.265-1.305} \rangle + \langle F_{\lambda=1.6-1.7} \rangle )} $ & continuum slope\\
            sKJ & (3) & $\frac{\langle F_{\lambda=1.265-1.305}\rangle - \langle F_{\lambda=2.12-2.16} \rangle}{0.5(\langle F_{\lambda=1.265-1.305} \rangle + \langle F_{\lambda=2.12-2.16} \rangle )} $ & continuum slope\\
            sH2O$^J$ & (3) & $\frac{\langle F_{\lambda=1.265-1.305}\rangle - \langle F_{\lambda=1.09-1.13} \rangle}{0.5(\langle F_{\lambda=1.265-1.305} \rangle + \langle F_{\lambda=1.09-1.13} \rangle )} $ & 1.1 $\mu$m H$_{2}$O absorption \\
            sH2O$^{H1}$ & (3) & $\frac{\langle F_{\lambda=1.6-1.7}\rangle - \langle F_{\lambda=1.45-1.48} \rangle}{0.5(\langle F_{\lambda=1.6-1.7} \rangle + \langle F_{\lambda=1.45-1.48} \rangle )} $ & $\sim$1.40 $\mu$m H$_{2}$O absorption\\
            sH2O$^{H2}$ & (3) & $\frac{\langle F_{\lambda=1.6-1.7}\rangle - \langle F_{\lambda=1.77-1.81} \rangle}{0.5(\langle F_{\lambda=1.6-1.7} \rangle + \langle F_{\lambda=1.77-1.81} \rangle )} $ & $\sim$1.85 $\mu$m H$_{2}$O absorption \\
            sH2O$^{K}$ & (3) & $\frac{\langle F_{\lambda=2.12-2.16}\rangle - \langle F_{\lambda=1.96-1.99} \rangle}{0.5(\langle F_{\lambda=2.21-2.16} \rangle + \langle F_{\lambda=1.96-1.99} \rangle )} $ & $\sim$2.4 $\mu$m H$_{2}$O absorption \\
            H$_{2}$O-1.2 & (4) & $\frac{\langle F_{\lambda=1.26-1.29}\rangle}{\langle F_{\lambda=1.13-1.16} \rangle} $ & $\sim$1.15 $\mu$m H$_{2}$O absorption\\
            H$_{2}$O-1.5 & (4) & $\frac{\langle F_{\lambda=1.57-1.59}\rangle}{\langle F_{\lambda=1.46-1.48} \rangle} $ & $\sim$1.4 $\mu$m H$_{2}$O absorption\\
            H$_{2}$O-2.0 & (4) & $\frac{\langle F_{\lambda=2.09-2.11}\rangle}{\langle F_{\lambda=1.975-1.995} \rangle} $ & $\sim$2.0 $\mu$m H$_{2}$O absorption\\
            CH$_{4}$-1.6 & (4) & $\frac{\langle F_{\lambda=1.56-1.60}\rangle}{\langle F_{\lambda=1.635-1.675} \rangle} $ & $\sim$1.6 $\mu$m CH$_{4}$ absorption\\
            CH$_{4}$-2.2 & (4) & $\frac{\langle F_{\lambda=2.08-2.12}\rangle}{\langle F_{\lambda=2.215-2.255} \rangle} $ & $\sim$2.2 $\mu$m CH$_{4}$ absorption\\
            J-FeH & (5) & $\frac{\langle F_{\lambda=1.195-1.205}\rangle}{\langle F_{\lambda=1.18-1.19} \rangle} $ & 1.2 $\mu$m FeH absorption\\
            H2OA & (5) & $\frac{\langle F_{\lambda=1.338-1.348}\rangle}{\langle F_{\lambda=1.308-1.318} \rangle} $ & $\sim$1.343 $\mu$m H$_{2}$O absorption \\
            H2OB & (5) & $\frac{\langle F_{\lambda=1.451-1.461}\rangle}{\langle F_{\lambda=1.565-1.575} \rangle} $ & $\sim$1.456 $\mu$m H$_{2}$O absorption\\
            H2OC & (5) & $\frac{\langle F_{\lambda=1.783-1.793}\rangle}{\langle F_{\lambda=1.717-1.727} \rangle} $ & $\sim$1.788 $\mu$m H$_{2}$O absorption\\
            H2OD & (5) & $\frac{\langle F_{\lambda=1.951-1.977}\rangle}{\langle F_{\lambda=2.062-2.088} \rangle} $ & $\sim$1.964 $\mu$m H$_{2}$O absorption\\
            CH$_{4}$A & (5) & $\frac{\langle F_{\lambda=1.725-1.735}\rangle}{\langle F_{\lambda=1.585-1.595}\rangle}$ & $\sim$1.730 $\mu$m CH$_{4}$ absorption\\
            CH$_{4}$B & (5) & $\frac{\langle F_{\lambda=2.195-2.205}\rangle}{\langle F_{\lambda=2.095-2.105}\rangle}$ & $\sim$2.2 $\mu$m CH$_{4}$ absorption\\
            CO & (5) & $\frac{\langle F_{\lambda=2.295-2.305}\rangle}{\langle F_{\lambda=2.280-2.290}\rangle}$ & $\sim$2.295 $\mu$m CO band\\
            H2O-1 & (6) & $\frac{\langle F_{\lambda=1.335-1.345}\rangle}{\langle F_{\lambda=1.295-1.305} \rangle} $ & $\sim$1.34 $\mu$m H$_{2}$O absorption\\
            H2O-2 & (6) & $\frac{\langle F_{\lambda=2.035-2.045}\rangle}{\langle F_{\lambda=2.145-2.155} \rangle} $ & $\sim$2.0 $\mu$m H$_{2}$O absorption\\
            FeH & (6) & $\frac{\langle F_{\lambda=1.1935-1.2065}\rangle}{\langle F_{\lambda=1.2235-1.2365} \rangle} $ & 1.2 $\mu$m FeH absorption\\
            H2O-K2 & (7) & $\frac{\langle F_{\lambda=2.07-2.09}\rangle / \langle F_{\lambda=2.235-2.255}\rangle}{\langle F_{\lambda=2.235-2.255}\rangle / \langle F_{\lambda=2.36-2.38}\rangle} $ & $\sim$2.07-2.38 $\mu$m H$_{2}$O absorption\\
            H2O-H & (8) & $\frac{\langle F_{\lambda=1.595-1615}\rangle / \langle F_{\lambda=1.68-1.7}\rangle}{\langle F_{\lambda=1.68-1.7}\rangle / \langle F_{\lambda=1.76-1.78}\rangle} $ & H-band H$_{2}$O absorption\\
            H2O-K & (8) & $\frac{\langle F_{\lambda=2.18-2.2}\rangle / \langle F_{\lambda=2.27-2.29}\rangle}{\langle F_{\lambda=2.27-2.29}\rangle / \langle F_{\lambda=2.36-2.38}\rangle} $ & K-band H$_{2}$O absorption\\
            Q & (9) & $\frac{\langle F_{\lambda=2.07-2.13}\rangle} {\langle F_{\lambda=2.267-2.285} \rangle} \left[ \frac{\langle F_{\lambda=2.4-2.5}\rangle}{\langle F_{\lambda=2.267-2.285} \rangle} \right] ^{1.22} $ & 1.7-2.1 $\mu$m and $>$2.4 $\mu$m H$_{2}$O absorption \\
            
            WH & (10) & $\frac{\langle F_{\lambda=1.552-1.572}\rangle}{\langle F_{\lambda=1.655-1.675} \rangle} $ & H$_{2}$O absorption\\
            WK & (10) & $\frac{\langle F_{\lambda=2.04-2.06}\rangle}{\langle F_{\lambda=2.18-2.2} \rangle} $ & H$_{2}$O absorption\\
            QH & (10) & $\frac{\langle F_{\lambda=1.552-1.572}\rangle}{\langle F_{\lambda=1.655-1.675} \rangle} \left[ \frac{\langle F_{\lambda=1.73-1.75}\rangle}{\langle F_{\lambda=1.655-1.675} \rangle} \right] ^{1.581} $ & H$_{2}$O absorption \\
            QK & (10) & $\frac{\langle F_{\lambda=2.04-2.06}\rangle}{\langle F_{\lambda=2.182-2.202} \rangle} \left[ \frac{\langle F_{\lambda=2.33-2.35}\rangle}{\langle F_{\lambda=2.182-2.202} \rangle} \right] ^{1.14} $ & H$_{2}$O absorption \\
            $\omega_{0}$ & (11) & $\left( \frac{W_{H2O}}{-0.0105}^*-\frac{W_{H2O-1}}{-0.0102}^* \right) \left( \frac{1}{-0.0105}+\frac{1}{-0.0102} \right)^{-1}$ & $\sim$1.50-1.57 $\mu$m H$_{2}$O absorption\\
            $\omega_{D}$ & (11) & $\left( \frac{W_{H2OD}}{0.0099}^*-\frac{W_{H2O-1}}{-0.0102} \right) \left( \frac{1}{0.0099}+\frac{1}{-0.0102} \right)^{-1}$ & $\sim$1.964 $\mu$m H$_{2}$O absorption\\
            $\omega_{2}$ & (11) & $\left( \frac{W_{H2O-2}}{0.0098}^*-\frac{W_{H2O-1}}{-0.0102} \right) \left( \frac{1}{0.0098}+\frac{1}{-0.0102} \right)^{-1}$ & $\sim$2.0 $\mu$m H$_{2}$O absorption\\
            K1 & (12) & $\frac{\langle F_{\lambda=2.10-2.18}\rangle - \langle F_{\lambda=1.96-2.04} \rangle}{0.5(\langle F_{\lambda=2.10-2.18} \rangle + \langle F_{\lambda=1.96-2.04} \rangle )} $ & 2.0 to 2.14 $\mu$m H$_{2}$O absorption \\
            K2 & (12) & $\frac{\langle F_{\lambda=2.20-2.28}\rangle - \langle F_{\lambda=2.10-2.18} \rangle}{0.5(\langle F_{\lambda=2.20-2.28} \rangle + \langle F_{\lambda=2.10-2.18} \rangle )} $ & 2.14 to 2.24 $\mu$m H$_{2}$ absorption \\
            H$_{2}$O-A & (13) & $\frac{\langle F_{1.12-1.17}\rangle}{ \langle F_{1.25-1.28}\rangle }$ & 1.15 $\mu$m H$_{2}$O/CH$_{4}$ \\
            H$_{2}$O-B & (13) & $\frac{{\langle}F_{1.505-1.525}{\rangle}}{{\langle}F_{1.575-1.595}{\rangle}}$ & 1.4  $\mu$m H$_2$O\\
            H$_{2}$O-C & (13) & $\frac{{\langle}F_{2.00-2.04}{\rangle}}{{\langle}F_{2.09-2.13}{\rangle}}$ & 1.9  $\mu$m H$_2$O \\
            CH$_{4}$-A & (13) & $\frac{{\langle}F_{1.295-1.325}{\rangle}}{{\langle}F_{1.25-1.28}{\rangle}}$ & 1.3  $\mu$m CH$_4$ \\
            CH$_{4}$-B & (13) & $\frac{{\langle}F_{1.64-1.70}{\rangle}}{{\langle}F_{1.575-1.595}{\rangle}}$ & 1.6  $\mu$m CH$_4$\\
            CH$_{4}$-C & (13) & $\frac{{\langle}F_{2.225-2.275}{\rangle}}{{\langle}F_{2.09-2.13}{\rangle}}$ & 2.2  $\mu$m CH$_4$\\
            H/J & (13) & $\frac{{\langle}F_{1.50-1.75}{\rangle}}{{\langle}F_{1.20-1.325}{\rangle}}$ & NIR color\\
    
            K/H & (13) & $\frac{{\langle}F_{2.00-2.30}{\rangle}}{{\langle}F_{1.50-1.75}{\rangle}}$ & NIR color \\
            CO & (13) & $\frac{{\langle}F_{2.325-2.375}{\rangle}}{{\langle}F_{2.09-2.13}{\rangle}}$ & 2.3  $\mu$m CO \\
            2.11/2.07 & (13) & $\frac{{\langle}F_{2.10-2.12}{\rangle}}{{\langle}F_{2.06-2.08}{\rangle}}$ & K-band shape/CIA H$_{2}$\\
            K shape & (13) & $\frac{\langle F_{2.12-2.13}\rangle - \langle F_{2.15-2.16}\rangle}{\langle F_{2.17-2.18}\rangle - \langle F_{2.19-2.20}\rangle}$ & K-band shape/CIA H$_{2}$\\
            H$_{2}$O-J & (14) & $\frac{{\langle}F_{1.14-1.165}{\rangle}}{{\langle}F_{1.26-1.285}{\rangle}}$ & 1.15 $\mu$m H$_{2}$O \\
            H$_{2}$O-H & (14) & $\frac{{\langle}F_{1.48-1.52}{\rangle}}{{\langle}F_{1.56-1.60}{\rangle}}$ & 1.40 $\mu$m H$_{2}$O \\
            H$_{2}$O-K & (14) & $\frac{{\langle}F_{1.975-1.995}{\rangle}}{{\langle}F_{2.08-2.10}{\rangle}}$ & 1.90 $\mu$m H$_{2}$O \\
            CH$_{4}$-J & (14) & $\frac{{\langle}F_{1.315-1.34}{\rangle}}{{\langle}F_{1.26-1.285}{\rangle}}$ & 1.32 $\mu$m CH$_{4}$ \\
            CH$_{4}$-H & (14) & $\frac{{\langle}F_{1.635-1.675}{\rangle}}{{\langle}F_{1.56-1.60}{\rangle}}$ & 1.65 $\mu$m CH$_{4}$ \\
            CH$_{4}$-K & (14) & $\frac{{\langle}F_{2.215-2.255}{\rangle}}{{\langle}F_{2.08-2.12}{\rangle}}$ & 2.20 $\mu$m CH$_{4}$ \\
            K/J & (14) & $\frac{{\langle}F_{2.060-2.10}{\rangle}}{{\langle}F_{1.25-1.29}{\rangle}}$ & J$-$K colour \\
            H-dip & (15) & $\frac{{\langle}F_{1.61-1.64}{\rangle}} {{\langle}F_{1.56-1.59}{\rangle} + {\langle}F_{1.66-1.69}{\rangle}}$ & 1.65$\mu$m CH$_{4}$ \\ 
\label{tab_app:spec_indices}
\end{longtable}

\begin{minipage}{0.9\textwidth}
\renewcommand{\footnoterule}{}
\tablefoot{(1) \citet{Allers_2007}; (2) \citet{AlmendrosAbad_2022}; (3) \citet{Testi_2001}; (4) \citet{Geballe_2002}; (5) \citet{McLean_2003}; (6) \citet{Slesnick_2004}; (7) \citet{Rojas_Ayala_2012}; (8) \citet{Covey_2010}; (9) \citet{Cushing_2000}; (10) \citet{Weights_2009}; (11) \citet{Zhang_2018}; (12) \citet{Tokunaga_1999}; (13) \citet{Burgasser_2002}; (14) \citet{Burgasser_2006}; (15) \citet{Burgasser_2010}.

* $W_{index}=-2.5\log{(index)}$, where index are the H2O \citep{Allers_2007}, H2O-D \citep{McLean_2003}, H2O-2 and H2O-1 \citep{Slesnick_2004} spectral indices.}
\end{minipage}
\twocolumn

\begin{table*}
\caption{Gravity-sensitive spectral indices inspected.}
	\begin{center}
        \begin{tabular}{l c c c}       
            \hline\hline
            Index & Reference & Formula & Feature Measured \\
            \hline
            FeH$_{J}$ & (1) & $\left(\frac{\lambda_{1.200}-\lambda_{1.192}}{\lambda_{1.208}-\lambda_{1.192}} F_{1.208} + \frac{\lambda_{1.208}-\lambda_{1.200}}{\lambda_{1.208}-\lambda_{1.192}} F_{1.192} \right) / F_{1.200}$ & 1.2 $\mu$m FeH absorption\\
            KI$_{J}$ & (1) & $\left(\frac{\lambda_{1.244}-\lambda_{1.220}}{\lambda_{1.270}-\lambda_{1.220}} F_{1.270} + \frac{\lambda_{1.270}-\lambda_{1.244}}{\lambda_{1.270}-\lambda_{1.220}} F_{1.220} \right) / F_{1.244}$ & 1.244, 1.253$\mu$m KI doublet\\
            H-cont & (1) & $\left(\frac{\lambda_{1.560}-\lambda_{1.470}}{\lambda_{1.670}-\lambda_{1.470}} F_{1.670} + \frac{\lambda_{1.670}-\lambda_{1.560}}{\lambda_{1.670}-\lambda_{1.470}} F_{1.470} \right) / F_{1.560}$ & H-band shape\\
            VO$_{z}$ & (1) & $\left(\frac{\lambda_{1.058}-\lambda_{1.035}}{\lambda_{1.087}-\lambda_{1.035}} F_{1.087} + \frac{\lambda_{1.087}-\lambda_{1.058}}{\lambda_{1.087}-\lambda_{1.035}} F_{1.035} \right) / F_{1.058}$ & 1.06 $\mu$m VO absorption\\
            TLI-g & (2) & $\frac{\langle F_{\lambda=1.56-1.58}\rangle}{\langle F_{\lambda=1.625-1.635} \rangle} $ & 1.5-1.565 $\mu$m H$_{2}$O absorption\\
            H$_{2}$(K) & (3) & $\frac{\langle F_{\lambda=2.16-2.18}\rangle}{\langle F_{\lambda=2.23-2.25} \rangle} $ & K-band peak\\
            HPI & (4) & $\frac{\langle F_{\lambda=1.675-1.685}\rangle}{\langle F_{\lambda=1.495-1.505} \rangle} $ & H-band peak for young objects\\
            \hline 
		\end{tabular}
		\tablefoot{(1) \citet{Allers_Liu_2013}; (2) \citet{AlmendrosAbad_2022}; (3) \citet{Canty_2013}; (4) \citet{scholz_2012}}
	\end{center}
	\label{tab_app:youth_indices}
\end{table*}

\end{appendix}

\end{document}